\documentclass[submission,copyright,creativecommons,sharealike,noncommercial]{eptcs}
\providecommand{\event}{QPL 2018}
\pdfoutput=1

\usepackage{mathtools} % Loads and extends amsmath
\usepackage{amssymb} % Extram mathematical symbols (loads amsfonts)
\usepackage{bbold} % Sans-serif blackboard bold font for numbers
\usepackage{amsthm} % Theorem environments
\usepackage{stmaryrd} % Some maths symbols for logic and computer science
\usepackage{relsize} % Additional relative sizes for fonts
\usepackage{microtype} % Improves appearance of writing
\usepackage{multicol} % Multi-column environments
\usepackage{csquotes} % Environments for quotes
\usepackage{hyperref} % Hyperlink citations, comment out for arxiv submission 
\usepackage[nocompress]{cite} % Comment out if using natbib or apacite
\usepackage{graphicx} % Import of graphics
\usepackage[usenames,dvipsnames]{xcolor} % Introduces colour names
\usepackage{tikz} % TikZ
\usepackage{circuitikz} % TikZ circuit diagrams 
\usetikzlibrary{
	arrows,
	shapes,
	decorations,
	intersections,
	backgrounds,
	positioning,
	circuits.ee.IEC
	}

		\numberwithin{theorem_c}{section} % ... numbered within sections
		\numberwithin{equation}{section} % Equations are also numbered within sections (but have a separate counter)
		\theoremstyle{plain}

		\newtheoremstyle{exampstyle}
		  {2mm} % Space above
		  {2mm} % Space below
		  {\itshape} % Body font
		  {} % Indent amount
		  {\bfseries} % Theorem head font
		  {.} % Punctuation after theorem head
		  {.5em} % Space after theorem head
		  {} % Theorem head spec (can be left empty, meaning `normal')

		\theoremstyle{exampstyle}

	\newcommand{\goodchi}{\protect\raisebox{2pt}{$\chi$}} % A chi letter raised at line level, for those instances where alignment is an issue
	\newcommand{\naturals}{\mathbb{N}} % Set of natural numbers
	\newcommand{\integers}{\mathbb{Z}} % Set of interer numbers
	\newcommand{\reals}{\mathbb{R}} % Set of real numbers
	\newcommand{\complexs}{\mathbb{C}} % Set of complex numbers
	\newcommand{\integersMod}[1]{\mathbb{Z}_{#1}} % Set/group/ring of integers mod #1
	\newcommand{\nonstd}[1]{\,^\star #1}
	\newcommand{\starNaturals}{\nonstd{\naturals}} % Set of non-standard naturals
	\newcommand{\starComplexs}{\nonstd{\complexs}} % Set of non-standard complex numbers
	\newcommand{\starReals}{\nonstd{\reals}} % Set of non-standard realsnumbers

	\newcommand{\suchthat}[2]{\left\{#1 \: \middle\vert \: #2\right\}} % Set of elements #1 such that condition #2 holds 
	\newcommand{\stdpartSym}{\operatorname{st}}
	\newcommand{\stdpart}[1]{\stdpartSym(#1)}
	
	\newcommand{\truncate}[1]{\bar{#1}}
	\newcommand{\liftSym}[1]{\operatorname{lift}_{#1}}

	\newcommand{\starIntegersMod}[1]{{\nonstd{\integersMod{#1}}}}
	\newcommand{\starIntegersModPow}[2]{{\nonstd{\integersMod{#1}^{#2}}}}

		\newcommand{\ket}[1]{\vert #1 \rangle} % Ket labelled #1
		\newcommand{\bra}[1]{\langle #1 \vert} % Bra labelled #1
		\newcommand{\braket}[2]{\langle #1 \vert #2 \rangle} % Inner product of bra labelled #1 with ket labelled #2
		
		\newcommand{\LtwoSym}{\operatorname{L}^2} % Symbol for L2 spaces
		\newcommand{\Ltwo}[1]{\LtwoSym[#1]} % L2 space over space #1
		\newcommand{\SpaceH}{\mathcal{H}} 
		\newcommand{\SpaceG}{\mathcal{G}}
		\newcommand{\SpaceK}{\mathcal{K}}

		\newcommand{\id}[1]{id_{#1}} % Identity morphism of object #1
		\newcommand{\Hom}[3]{\operatorname{Hom}_{\,#1}\left[#2,#3\right]} % Set of morphisms in category #1 from object #2 to object #3
		\newcommand{\HilbCategory}{\operatorname{Hilb}} % Category of Hilbert spaces
		\newcommand{\sHilbCategory}{\operatorname{sHilb}} % Category of separable Hilbert spaces
		\newcommand{\starHilbCategory}{^\star\!\HilbCategory} % Category of omega-truncated non-standard hilbert spaces
		
		\newcommand{\starsHilbCategory}{\nonstd{\operatorname{\sHilbCategory}}}
		\newcommand{\starsHilbCategoryNearStd}{\starsHilbCategory^{(std)}}

	\newcommand{\Dcolour}{black!80}
	\newcommand{\Xbwcolour}{black!80}
	\newcommand{\Zbwcolour}{white}
	\newcommand{\Ybwcolour}{black!15}
	\newcommand{\Wbwcolour}{black!50}
	\tikzset{
	  rectangle with rounded corners north west/.initial=4pt,
	  rectangle with rounded corners south west/.initial=4pt,
	  rectangle with rounded corners north east/.initial=4pt,
	  rectangle with rounded corners south east/.initial=4pt,
	}
	\makeatletter
	\pgfdeclareshape{rectangle with rounded corners}{
	  \inheritsavedanchors[from=rectangle] % this is nearly a rectangle
	  \inheritanchorborder[from=rectangle]
	  \inheritanchor[from=rectangle]{center}
	  \inheritanchor[from=rectangle]{north}
	  \inheritanchor[from=rectangle]{south}
	  \inheritanchor[from=rectangle]{west}
	  \inheritanchor[from=rectangle]{east}
	  \inheritanchor[from=rectangle]{north east}
	  \inheritanchor[from=rectangle]{south east}
	  \inheritanchor[from=rectangle]{north west}
	  \inheritanchor[from=rectangle]{south west}
	  \backgroundpath{% this is new
	    \southwest \pgf@xa=\pgf@x \pgf@ya=\pgf@y
	    \northeast \pgf@xb=\pgf@x \pgf@yb=\pgf@y
	    \pgfkeysgetvalue{/tikz/rectangle with rounded corners north west}{\pgf@rectc}
	    \pgfsetcornersarced{\pgfpoint{\pgf@rectc}{\pgf@rectc}}
	    \pgfpathmoveto{\pgfpoint{\pgf@xa}{\pgf@ya}}
	    \pgfpathlineto{\pgfpoint{\pgf@xa}{\pgf@yb}}
	    \pgfkeysgetvalue{/tikz/rectangle with rounded corners north east}{\pgf@rectc}
	    \pgfsetcornersarced{\pgfpoint{\pgf@rectc}{\pgf@rectc}}
	    \pgfpathlineto{\pgfpoint{\pgf@xb}{\pgf@yb}}
	    \pgfkeysgetvalue{/tikz/rectangle with rounded corners south east}{\pgf@rectc}
	    \pgfsetcornersarced{\pgfpoint{\pgf@rectc}{\pgf@rectc}}
	    \pgfpathlineto{\pgfpoint{\pgf@xb}{\pgf@ya}}
	    \pgfkeysgetvalue{/tikz/rectangle with rounded corners south west}{\pgf@rectc}
	    \pgfsetcornersarced{\pgfpoint{\pgf@rectc}{\pgf@rectc}}
	    \pgfpathclose
	 }
	}
	\makeatother

	\tikzset{->-/.style={decoration={markings,mark=at position #1 with {\arrow{>}}},postaction={decorate}}}
	\tikzset{-<-/.style={decoration={markings,mark=at position #1 with {\arrow{<}}},postaction={decorate}}}

	\tikzstyle{every picture}=[baseline=-0.25em,scale=0.5]
	\pgfdeclarelayer{edgelayer}
	\pgfdeclarelayer{nodelayer}
	\pgfsetlayers{edgelayer,nodelayer,main}

	\tikzstyle{box} = [draw,shape=rectangle,inner sep=2pt,minimum height=6mm,minimum width=6mm,fill=white] 
	\tikzstyle{boxlarge} = [draw,shape=rectangle,inner sep=2pt,minimum height=1.5cm,minimum width=8mm,fill=white] 
	\tikzstyle{boxLarge} = [draw,shape=rectangle,inner sep=2pt,minimum height=2cm,minimum width=10mm,fill=white] 
	\tikzstyle{boxsmall} = [draw,shape=rectangle,inner sep=2pt,minimum height=3mm,minimum width=3mm,fill=white] % small in all directions. Might one day use boxnarrow for small in the largeness direction only.
	\tikzstyle{dot} = [inner sep=0mm,minimum width=3mm,minimum height=3mm,draw,shape=circle,text depth=-0.1mm]
	\tikzstyle{Zbwdot} = [dot, fill=\Zbwcolour]
	\tikzstyle{Xbwdot} = [dot, fill=\Xbwcolour]
	\tikzstyle{Ybwdot} = [dot, fill=\Ybwcolour]
	\tikzstyle{Wbwdot} = [dot, fill=\Wbwcolour]
	\tikzstyle{antipode} = [boxsmall] 

	\tikzstyle{state} = [draw, rectangle with rounded corners,
	  rectangle with rounded corners north west=8pt,
	  rectangle with rounded corners south west=8pt,
	  rectangle with rounded corners north east=0pt,
	  rectangle with rounded corners south east=0pt,
	,inner sep=2pt,minimum height=6mm,minimum width=6mm,fill=white]
	\tikzstyle{statelarge} = [draw, rectangle with rounded corners,
	  rectangle with rounded corners north west=8pt,
	  rectangle with rounded corners south west=8pt,
	  rectangle with rounded corners north east=0pt,
	  rectangle with rounded corners south east=0pt,
	,inner sep=2pt,minimum height=1.5cm,minimum width=8mm,fill=white]
	\tikzstyle{stateLarge} = [draw, rectangle with rounded corners,
	  rectangle with rounded corners north west=8pt,
	  rectangle with rounded corners south west=8pt,
	  rectangle with rounded corners north east=0pt,
	  rectangle with rounded corners south east=0pt,
	,inner sep=2pt,minimum height=2cm,minimum width=8mm,fill=white]
	\tikzstyle{effect} = [draw, rectangle with rounded corners,
	  rectangle with rounded corners north west=0pt,
	  rectangle with rounded corners south west=0pt,
	  rectangle with rounded corners north east=8pt,
	  rectangle with rounded corners south east=8pt,
	,inner sep=2pt,minimum height=6mm,minimum width=6mm,fill=white]
	\tikzstyle{scalar}=[diamond,draw,inner sep=1pt,font=\small,fill=white]

	\tikzstyle{cdnode}=[fill=white]
	\tikzstyle{labelnode}=[fill=white]
	\tikzstyle{tightlabelnode}=[fill=white,inner sep = 0.1mm]
	\tikzstyle{none}=[inner sep=0pt]
	\tikzstyle{whiteline}=[-, line width=4pt, draw=white]
	\tikzstyle{trace}=[circuit ee IEC,thick,ground,scale=2.5]
	\tikzstyle{cotrace}=[circuit ee IEC,thick,ground,rotate=180,scale=2.5]
	\tikzstyle{upground}=[circuit ee IEC,thick,ground,rotate=90,scale=2.5]
	\tikzstyle{downground}=[circuit ee IEC,thick,ground,rotate=-90,scale=2.5]

	\tikzstyle{doubled} = [line width=1.8pt] % [line width=1.6pt] % [very thick]
	
	\tikzstyle{empty diagram}=[draw=gray!40!white,dashed,shape=rectangle,minimum width=1cm,minimum height=1cm]

\definecolor{red}{RGB}{255,128,128}

\title{Quantum Field Theory \\ in Categorical Quantum Mechanics}
\author{
	Stefano Gogioso\\
	University of Oxford \\
	\texttt{stefano.gogioso@cs.ox.ac.uk}
	\and
	Fabrizio Genovese \\
	University of Oxford \\
	\texttt{fabrizio.genovese@cs.ox.ac.uk}
}

\def\titlerunning{Quantum Field Theory in Categorical Quantum Mechanics}
\def\authorrunning{S. Gogioso \& F. Genovese}

\begin{document}

\maketitle

\begin{abstract}
	We use tools from non-standard analysis to formulate the building blocks of quantum field theory within the framework of categorical quantum mechanics.
	Building upon previous work, we construct an object of $\starHilbCategory$ having quantum fields as states and we show that the usual ladder and field operators can be defined as suitable endomorphisms. We deal with relativistic normalisation and we obtain the Lorentz invariant Heisenberg picture operators. By moving to a coherent perspective---where the classical time and momentum parameters are replaced by wavefunctions over the parameter spaces---we show that ladder operators and field operators can be obtained by applying the same morphism to plane waves and delta functions respectively. Finally, we formulate the commutation relations diagrammatically and we use them to derive the propagator.
\end{abstract}

% \vspace{-4mm}
\section{Introduction} 
\label{section_introduction}

One of the very first obstacles in the passage from quantum theory to quantum field theory is the lack of a suitable Hilbert space having quantum fields as vectors. In the simplest of cases---that of the scalar quantum fields---this is due to the commutation relations 
%$[a({\underline{p}}),a^\dagger(\underline{q})] = \delta(p-q) \id{}$ 
between ladder operators, that imply that the vectors 
%$\ket{\underline{p}} := a^\dagger(\underline{q})$ 
for single-particle states would have infinite square norm.
% \begin{equation}
% 	\braket{\underline{p}}{\underline{p}} = \bra{vacuum}a({\underline{p}})a^\dagger({\underline{p}})\ket{vacuum} = \delta(0)	
% \end{equation}
In order to deal with this obstacle, presentations of quantum field theory (we will broadly follow \cite{PeskinSchroeder,TongQFT}) typically work with operators and operator-valued distributions in place of vectors, obtaining amplitudes by sandwiching operators with the vacuum vector and jumping through hoops to avoid infinities.  

When using non-standard analysis \cite{Robinson1974}, those infinities are no longer an issue and it is perfectly sensible to treat quantum fields as vectors in the infinite tensor product of simple harmonic oscillators over momentum space. That is the approach taken here, building upon previous work on the non-standard approach  \cite{GogiosoGenovese2016,GogiosoGenovese2017} to infinite dimensional categorical quantum mechanics \cite{AbramskyCoecke2004,PicturingQuantumProcesses,AbramskyHeunen2012,HeunenReyes2017}. 

In Section \ref{section_recap}, we open with a brief summary of the non-standard approach to infinite dimensional categorical quantum mechanics, including the construction of the infinite tensor product of Hilbert spaces over momentum space, home to our formulation of quantum fields. 

In Section \ref{section_QFT}, we introduce the basic ingredients of quantum field theory within our framework. We quantise the Klein-Gordon equation by considering a field of simple harmonic oscillators over momentum space, over which we define ladder operators, quantum field states and field operators. We tackle the issue of relativistic normalisation of on-shell states and we define Lorentz-invariant operators in the Heisenberg picture. We interpret ladder/field operators as classically controlled by momentum, position and time parameters: we turn them into a single operator coherently controlled by wavefunctions over the parameter spaces, and we show that the original operators can be recovered by application to momentum, position and time eigenstates. Finally, we provide a diagrammatic formulation of the commutation relations in the Schr\"{o}dinger picture, and we use them to derive the commutation relations for the Heisenberg picture and a diagrammatic expression for the propagator. 

\newpage
\section{Brief recap of the story so far}
\label{section_recap}

\subsection{The dagger compact category \texorpdfstring{$\starHilbCategory$}{Star Hilb}}

The objects of $\starHilbCategory$ take the form of pairs $\SpaceH := (|\SpaceH|, P_\SpaceH)$, where $|\SpaceH|$ is a non-standard Hilbert space (the \textbf{underlying Hilbert space}) and $P_\SpaceH : |\SpaceH| \rightarrow |\SpaceH|$ is an internal non-standard $\starComplexs$-linear map, its \textbf{truncating projector}:
\begin{equation}
	P_\SpaceH = \sum_{n=1}^{D} \ket{e_n}\bra{e_n}
\end{equation}
where $\ket{e_n}_{n=1}^{D}$ is some family of orthonormal vectors in $|\SpaceH|$, for some $D \in \starNaturals$. The number $D$ is independent of the choice of family by Transfer Theorem, and can be used to consistently define the \textbf{dimension} of $\SpaceH$ to be $\dim{\SpaceH} := D \in \starNaturals$. The morphisms of $\starHilbCategory$ are internal non-standard $\starComplexs$-linear maps taking the following form, with the truncating projectors acting as identities:
\begin{equation}
	\Hom{\starHilbCategory}{\SpaceH}{\SpaceG} := \suchthat{\;P_\SpaceG \circ F \circ P_\SpaceH\;}{\;F:  |\SpaceH| \,\rightarrow\, |\SpaceG| \text{ internal linear map}}.
\end{equation}
The category $\starHilbCategory$ is a full subcategory of the Karoubi envelope for the category of non-standard Hilbert spaces and non-standard $\starComplexs$-linear maps. Morphisms can be represented as matrices by choosing orthonormal families of vectors which diagonalise the relevant truncating projectors:
\begin{equation}
	\truncate{F} := P_\SpaceG \circ F \circ P_\SpaceH = 
	\sum_{m=1}^{\dim{\SpaceG}} \sum_{n=1}^{\dim{\SpaceH}} 
	\ket{f_{m}} \Big( \bra{f_{m}} F \ket{e_{n}} \Big) \bra{e_{n}}.
\end{equation}
The tensor product, the symmetry isomorphisms, the dagger, the compact closed structure and the dagger biproducts can be defined as usual by looking at the matrix decomposition, and by Transfer Theorem they are invariant under different choices of diagonalising orthonormal sets. Similarly, unital special commutative $\dagger$-Frobenius algebras can be constructed for all orthonormal bases of an object $\SpaceH$ (i.e. for all orthonormal families diagonalising the truncating projector $P_\SpaceH$).

\subsection{Relation to standard separable Hilbert spaces}

Let $\sHilbCategory$ be the $\dagger$-SMC of standard separable Hilbert spaces and bounded linear maps. Let $\starsHilbCategory$ be the full subcategory of $\starHilbCategory$ given by those objects $\SpaceH$ such that $|\SpaceH| = \nonstd{V}$ for some separable standard Hilbert space $V$ and such that the truncating projector spans all near-standard vectors. Let $\starsHilbCategoryNearStd$ be the sub-$\dagger$-SMC of $\starsHilbCategory$ given by only considering near-standard morphisms. 

We can define a \textbf{standard part functor} $\stdpartSym : \; \starsHilbCategoryNearStd \rightarrow \sHilbCategory$, which acts as $\SpaceH \mapsto |\SpaceH|$ on objects and as $\truncate{F} \mapsto \stdpart{\truncate{F}}$ on morphisms. The standard part functor is $\complexs$-linear, and identifies two near-standard maps $\truncate{F},\truncate{G}: \SpaceH \rightarrow \SpaceK$ if and only if $\truncate{F}-\truncate{G}$ has infinitesimal operator norm; this defines an equivalence relation on morphisms in $\starsHilbCategoryNearStd$, which we denote by $\sim$ and refer to as \textbf{infinitesimal equivalence}. For each infinite natural $\omega$, we can define a weak \textbf{truncation functor} $\liftSym{\omega}: \sHilbCategory \rightarrow\, \starsHilbCategoryNearStd$, which acts as $V \mapsto (V,P^{(V)})$ on objects and sends the standard morphism $f: V \rightarrow W $ to the non-standard morphism $\truncate{F} := P^{(W)} \circ F \circ P^{(V)}$ (here $F := \nonstd{\!f}$ is the non-standard extension of $f$). 

Theorem 2.1 from \cite{GogiosoGenovese2017} states that $\stdpartSym$ and $\liftSym{\omega}$ define a weak equivalence between $\sHilbCategory$ and $\starsHilbCategoryNearStd_{\omega}$, the full subcategory of $\starsHilbCategoryNearStd$ spanned by those objects $\SpaceH$ having dimension $\dim{\SpaceH} \in \starNaturals$ which is either a finite natural or the infinite natural $\omega$. The essence of Theorem 2.1 is that $\sHilbCategory$ is equivalent to the subcategory $\starsHilbCategoryNearStd_{\omega}$, as long as we take care to equate morphisms which are infinitesimally close. The equivalence allows one to prove results about $\sHilbCategory$ by working in $\starHilbCategory$ and taking advantage of the full CQM machinery, according to the following general recipe:
\vspace{-2mm}
\begin{multicols}{2}
\begin{enumerate}
	\itemsep0em
	\item[(i)] start from a morphism in $\sHilbCategory$; 
	\item[(ii)] lift to $\starsHilbCategoryNearStd_{\omega}$ via the lifting functor;
	\item[(iii)] work in $\starHilbCategory$, obtain a result in $\starsHilbCategoryNearStd_{\omega}$; 
	\item[(iv)] descend to $\sHilbCategory$ via the standard part functor. 
\end{enumerate}
\end{multicols}

\subsection{The non-standard cyclic group \texorpdfstring{$\starIntegersMod{2 \omega+1}$}{of integers modulo infinities}}

The abelian group $\starIntegersMod{2 \omega+1}$ is defined to be the internal set of non-standard integers $\{-\omega,...,+\omega\}$ endowed with $0$ as unit and with addition $\oplus$ modulo $2\omega+1$ as group operation. The group $\starIntegersMod{2 \omega + 1}$ has the integers as a subgroup: if $k,h \in \integers$ are standard integers, then certainly $-\omega \leq k+h \leq +\omega$, and hence $k \oplus h = k+h$. In this work, we will be interested in using the group  $(\starIntegersModPow{2\omega+1}{n},\oplus,\underline{0})$ to approximate $n$-dimensional real space $\reals^n$, by using a non-standard lattice of infinitesimal mesh and working up to infinitesimal equivalence. Concretely, this is done by considering the embedding of  $\starIntegersModPow{2 \omega + 1}{n}$ into $\starReals^n$ as the lattice $\frac{1}{\omega_{ir}}\starIntegersModPow{2 \omega + 1}{n}$, where we assume that the infinite natural $\omega$ has been chosen in such a way that we can write $2\omega+1 = \omega_{ir}\omega_{uv}$ for some odd infinite naturals $\omega_{ir},\omega_{uv} \in \starNaturals^+$ 
	\footnote{Note that the convention here is different from the $\omega = \omega_{uv} \omega_{ir}$ originally adopted in \cite{GogiosoGenovese2017}.} 
	\footnote{The subscripts ``ir'' and ``uv'' stand for ``infrared'' and ``ultraviolet'' respectively. They are a reference to the infrared and ultraviolet cut-offs used as part of certain renormalisation techniques in QFT.}
	\footnote{Alternatively, we can first choose odd infinite naturals $\omega_{ir},\omega_{uv}$ and subsequently define $\omega:=\frac{\omega_{ir}\omega_{uv}-1}{2}$.}
and we send $\underline{k} \in\starIntegersModPow{2 \omega + 1}{n}$ to $\underline{p}:= \underline{k}/\omega_{ir} \in \starReals^n$. The standard $n$-dimensional real space $\reals^n$ is recovered from the non-standard lattice $\frac{1}{\omega_{ir}}\starIntegersModPow{2 \omega + 1}{n}$ by restricting our attention to the subgroup of those $\underline{k} \in \starIntegersModPow{2\omega+1}{n}$ such that $\underline{p} := \underline{k}/\omega_{ir}$ is a near-standard vector 
	\footnote{One has to be careful to impose this restriction externally to the non-standard model---by hand, if you will---because the aforementioned subgroup does not correspond to an internal subset of $\starIntegersModPow{2\omega+1}{n}$.} 
in $\starReals^n$, and then quotienting by infinitesimal equivalence of vectors to obtain the group $\reals^n$. In this sense, the non-standard lattice $\frac{1}{\omega_{ir}}\starIntegersModPow{2\omega+1}{n}$ in $\starReals^n$ approximates the real $n$-dimensional space $\reals^n$ to infinitesimal mesh~$\frac{1}{\omega_{ir}}$, covering it all the way up to some infinity $\omega_{uv}$ where the lattice circles around.

By using the non-standard complex group algebra for the abelian group $\frac{1}{\omega_{ir}}\starIntegersModPow{2\omega+1}{n}$, we can construct an object of $\starHilbCategory$ that can be used to deal with quantum particles living in $n$-dimensional real space, corresponding to the standard Hilbert space $\Ltwo{\reals^n}$. We consider the following orthogonal family of vectors in $\nonstd{\reals^n} \rightarrow \starComplexs$, indexed by $\underline{p} \in \frac{1}{\omega_{ir}}\starIntegersModPow{2\omega+1}{n}$ and having square norm $\omega_{ir}^n$:
\begin{equation}
	\ket{\goodchi_{\underline{p}}} := \underline{x} \mapsto e^{i 2\pi \, \underline{p}\cdot \underline{x}}
\end{equation} 
We use this family to define a $\big((2 \omega + 1)^n\big)$-dim object $\starComplexs[\frac{1}{\omega_{ir}} \starIntegersModPow{2\omega+1}{n}] := (\nonstd{\Ltwo{\reals^n}},P_{\starComplexs[\frac{1}{\omega_{ir}} \starIntegersModPow{2\omega+1}{n}]})$ of $\starHilbCategory$, by considering the following truncating projector:
\begin{equation}
	P_{\starComplexs[\frac{1}{\omega_{ir}} \starIntegersModPow{2\omega+1}{n}]} 
	:= 
	\sum_{\underline{p} \in  \frac{1}{\omega_{ir}}\starIntegersModPow{2\omega+1}{n}} 
	\!\!\!\!\frac{1}{\omega_{ir}^n}\;\;
	\ket{\goodchi_{\underline{p}}}\bra{\goodchi_{\underline{p}}}
\end{equation}  
We refer to the orthogonal basis $\ket{\goodchi_{\underline{p}}}_{\underline{p} \in  \frac{1}{\omega_{ir}}\starIntegersModPow{2\omega+1}{n}}$ as the \textbf{momentum eigenstates} for $\starComplexs[\frac{1}{\omega_{ir}} \starIntegersModPow{2\omega+1}{n}]$, and we can use it to construct a basis of \textbf{position eigenstates} as the following orthogonal family, indexed by $\underline{x} \in \frac{1}{\omega_{uv}}\starIntegersModPow{2\omega+1}{n}$ and with square norm $\omega_{uv}^n$:
\begin{equation}
	\ket{\delta_{\underline{x}}} 
	:= 
	\sum_{\underline{p} \in  \frac{1}{\omega_{ir}}\starIntegersModPow{2\omega+1}{n}} 
	\!\!\!\!\frac{1}{\omega_{ir}^n}\;\;
	e^{-i 2\pi \, \underline{p}\cdot \underline{x}} \ket{\goodchi_{\underline{p}}}
\end{equation}
We can check that these truly behave as position/momentum eigenstates by computing their inner product:
\begin{equation}
	\braket{\delta_{\underline{x}}}{\goodchi_{\underline{p}}} = \frac{1}{\omega_{ir}^n}  e^{i2\pi \,(\underline{p}\cdot\underline{x})} \braket{\goodchi_{\underline{p}}}{\goodchi_{\underline{p}}} = e^{i2\pi \,\underline{p}\cdot\underline{x}} = \goodchi_{\underline{p}}(\underline{x})
\end{equation}
It is shown in \cite{GogiosoGenovese2017,CategoricalQuantumDynamics} that these bases correspond to a strongly complementary pair $(\hbox{\begin{tikzpicture} [scale=1.2,transform shape] %% DO NOT CHANGE

\def\deltax{0.3} %% CAN BE CHANGED
\def\deltay{0.5} %% DO NOT CHANGE

%\path[use as bounding box] (-\deltax,-\deltay) rectangle (\deltax,\deltay);

\node [dot, fill=\Zbwcolour] (mult) at (0,0) {};
%\node (mult_label_out) at (0,+\deltay) {};
%\draw[-] (mult) to (mult_label_out);

%\draw (current bounding box.south west) rectangle (current bounding box.north east);
\end{tikzpicture}
}\!\!,\hbox{\begin{tikzpicture} [scale=1.2,transform shape] %% DO NOT CHANGE

\def\deltax{0.3} %% CAN BE CHANGED
\def\deltay{0.5} %% DO NOT CHANGE

%\path[use as bounding box] (-\deltax,-\deltay) rectangle (\deltax,\deltay);

\node [dot, fill=\Dcolour] (mult) at (0,0) {};
%\node (mult_label_out) at (0,+\deltay) {};
%\draw[-] (mult) to (mult_label_out);

%\draw (current bounding box.south west) rectangle (current bounding box.north east);
\end{tikzpicture}
}\!\!)$ of quasi-special commutative dagger Frobenius algebras\cite{CoeckeDuncan2011}---the \textbf{momentum observable} $\hbox{\begin{tikzpicture} [scale=1.2,transform shape] %% DO NOT CHANGE

\def\deltax{0.3} %% CAN BE CHANGED
\def\deltay{0.5} %% DO NOT CHANGE

%\path[use as bounding box] (-\deltax,-\deltay) rectangle (\deltax,\deltay);

\node [dot, fill=\Zbwcolour] (mult) at (0,0) {};
%\node (mult_label_out) at (0,+\deltay) {};
%\draw[-] (mult) to (mult_label_out);

%\draw (current bounding box.south west) rectangle (current bounding box.north east);
\end{tikzpicture}
}\!\!$ and the \textbf{position observable} $\hbox{\begin{tikzpicture} [scale=1.2,transform shape] %% DO NOT CHANGE

\def\deltax{0.3} %% CAN BE CHANGED
\def\deltay{0.5} %% DO NOT CHANGE

%\path[use as bounding box] (-\deltax,-\deltay) rectangle (\deltax,\deltay);

\node [dot, fill=\Dcolour] (mult) at (0,0) {};
%\node (mult_label_out) at (0,+\deltay) {};
%\draw[-] (mult) to (mult_label_out);

%\draw (current bounding box.south west) rectangle (current bounding box.north east);
\end{tikzpicture}
}\!\!$---with strong complementarity playing the role of position-momentum duality in the CQM framework.

\subsection{Quantum fields in real space}

Consider an object $\SpaceH:=(|\SpaceH|,P_\SpaceH)$ of $\starHilbCategory$, with truncating projector $P_\SpaceH $ decomposed as $P_\SpaceH = \sum_{d=1}^{D} \ket{e_d}\bra{e_d}$ for some orthonormal family $\ket{e_d}_{d=1}^{D}$ in $|\SpaceH|$. We wish to construct the infinite tensor product $\SpaceH^{\otimes\frac{1}{\omega_{ir}}\starIntegersModPow{2\omega+1}{n}}$, which will allow us to model $\SpaceH$-valued quantum fields living on the lattice $\frac{1}{\omega_{ir}}\starIntegersModPow{2 \omega + 1}{n}$. Upon near-standard restriction and quotient by infinitesimal equivalence, this allows us to  deal with standard quantum fields on $n$-dimensional real space $\reals^n$. 

Define the shorthands $D := \dim{\SpaceH}$ and $\mu := (2 \omega + 1)^n$, and consider the following orthonormal family $\ket{e_{\underline{s}}}$ of non-standard states in $|\SpaceH|^{\otimes \mu}$, indexed by strings $\underline{s} \in \{1,...,D\}^{\mu}$:
\begin{equation}\label{eqn_latticeFieldBasis}
	\ket{e_{\underline{s}}} := \bigotimes_{k_1 = -\omega}^{+\omega} ... \bigotimes_{k_n=-\omega}^{+ \omega} \ket{e_{\underline{s}(\underline{k})}}
\end{equation}
The definition in terms of $\underline{k} \in \starIntegersModPow{2\omega+1}{n}$ is the clearest from a mathematical standpoint, but a definition in terms of $\underline{p} := \underline{k}/\omega_{ir} \in \frac{1}{\omega_{ir}}\starIntegersModPow{2\omega+1}{n}$ carries more immediate physical significance. As a consequence, this work will use the notations $\otimes_{p = -\omega_{uv}}^{+\omega_{uv}}$ and  $\sum_{p = -\omega_{uv}}^{+\omega_{uv}}$ in place of $\otimes_{k = -\omega}^{+\omega}$ and $\sum_{k = -\omega}^{+\omega}$ respectively, as well as $\underline{s}(\underline{p})$ in place of $\underline{s}(\underline{k})$ for $\underline{k} := \omega_{ir}\underline{p}$. 
Hence the orthonormal family chosen above equivalently reads:
\begin{equation}\label{eqn_latticeFieldBasisP}
	\ket{e_{\underline{p}}} := \bigotimes_{p_1 = -\omega_{uv}}^{+\omega_{uv}} ... \bigotimes_{p_n=-\omega_{uv}}^{+ \omega_{uv}} \ket{e_{\underline{s}(\underline{p})}}
\end{equation}
To model $\SpaceH$-valued quantum fields on the lattice $\frac{1}{\omega_{ir}}\starIntegersModPow{2\omega+1}{n}$ within our framework, we define the object $\SpaceH^{\otimes\frac{1}{\omega_{ir}}\starIntegersModPow{2\omega+1}{n}} := (|\SpaceH|^{\otimes\mu}, P_{ \SpaceH^{\otimes\frac{1}{\omega_{ir}}\starIntegersModPow{2\omega+1}{n}}})$ of $\starHilbCategory$, where the truncating projector is defined by:
\begin{equation}
	P_{ \SpaceH^{\otimes\frac{1}{\omega_{ir}}\starIntegersModPow{2\omega+1}{n}}} = \sum_{\underline{s}} \ket{e_{\underline{s}}}\bra{e_{\underline{s}}} 
\end{equation} 
In \cite{GogiosoGenovese2017}, it is shown that the $\SpaceH^{\otimes\frac{1}{\omega_{ir}}\starIntegersModPow{2\omega+1}{n}}$ behaves as a genuine tensor product---at least in the case, of interest here, where $\SpaceH = \nonstd{V}$ arises from a separable standard Hilbert space $V$---in the sense that it satisfies a universal property that allows multilinear maps $\tilde{F}:\prod_{\underline{p} \in \frac{1}{\omega_{ir}}\starIntegersModPow{2\omega+1}{n}} \SpaceH \longrightarrow \SpaceK$---where $\SpaceK$ with $|\SpaceK| = \nonstd{W}$ for some separable standard Hilbert space $W$---to be lifted uniquely to linear maps $F:\SpaceH^{\otimes\frac{1}{\omega_{ir}}\starIntegersModPow{2\omega+1}{n}} \longrightarrow \SpaceK$. It is also shown that $\prod_{\underline{p} \in \frac{1}{\omega_{ir}}\starIntegersModPow{2\omega+1}{n}} \SpaceH$ is related to the direct integral $\int_{\reals^n}^{\oplus} \,V \,d\underline{p}$ \cite{vonNeumann1939,vonNeumann1949} by restriction and quotient: any continuous $\varphi: \reals^n \rightarrow V$ can be lifted to the near-standard $\nonstd{\varphi}: \frac{1}{\omega_{ir}}\starIntegersModPow{2\omega+1}{n} \rightarrow \nonstd{V}$, and any such $\varphi$ can vice-versa be reconstructed from $\nonstd{\varphi}$ by setting $\varphi(\underline{q}):=\stdpart{\nonstd{\varphi}(\underline{p})}$ for any $\underline{p} \in \frac{1}{\omega_{ir}}\starIntegersModPow{2\omega+1}{n}$ such that $\stdpart{\underline{p}} = \underline{q} \in \reals^n$. The relationship to $\int_{\reals^n}^{\oplus} \,V \,d\underline{p}$ together with the universal property shows how $\SpaceH^{\otimes\frac{1}{\omega_{ir}}\starIntegersModPow{2\omega+1}{n}}$ can be used to work with standard quantum fields.

\newpage
\section{Quantum Field Theory}
\label{section_QFT}

As part of canonical quantisation, classical fields from the Lagrangian formalism are translated into certain operator-valued distributions, also known as \textit{field operators}, acting upon quantum states living in a Fock space. Using the field operators, the classical Lagrangian can be translated into the dynamics and interactions of the quantum field theory, so it is no surprise that they occupy the vast majority of the literature dedicated to the subject. 

It is worth noting, however, that the field operators play a very different role from the classical fields that they originally quantised: classical fields \textit{are} states of a classical system, while field operators \textit{act upon} states of a quantum system (e.g. the vacuum). In this sense, the closest correspondents in quantum field theory to the fields of classical field theory or the wavefunctions of quantum mechanics are, in fact, the quantum states in the Fock space. Just as $\complexs^2$ is the space of quantum states for a qubit, so the Fock space is the space of quantum states for a quantum field. And just as we freely refer to the object $\complexs^2$ as a qubit, so we take the liberty to refer to the Fock space as a \textbf{quantum field}. We will use the term \textbf{field operator} when talking about the operator-valued distributions obtained by canonical quantisation.

\subsection{Simple Harmonic Oscillator}

Picking things up from the very last section of \cite{GogiosoGenovese2017}, let's consider the textbook example of the real scalar field, a relativistic classical field $\phi(\underline{x},t)$ satisfying the \textbf{Klein-Gordon equation}:
\begin{equation}
	\partial_\mu \partial^\mu \phi + \left(\frac{mc}{\hbar}\right)^2 \phi = 0
\end{equation}
When looking at the field in momentum space $\phi(\underline{p},t)$, the Klein-Gordon equation becomes:
\begin{equation}
	\Big(\hbar^2\frac{\partial^2}{\partial t^2} + (|\underline{p}|c)^2+(mc^2)^2\Big) \phi(\underline{p},t) = 0
\end{equation}
Hence a momentum space solution $\phi(\underline{p},t)$ to the Klein-Gordon equation can be thought of as a field of simple harmonic oscillators, each oscillator vibrating with its own amplitude and at a frequency $\nu_{\underline{p}} = \frac{1}{\hbar}E_{\underline{p}}$ for each point $\underline{p} \in \reals^3$ of momentum space, where the energy $E_{\underline{p}}$ is given by the relativistic dispersion relation:
\begin{equation}
	 E_{\underline{p}} := \sqrt{(|\underline{p}|c)^2+(mc^2)^2} 
\end{equation}
In order to quantise the real scalar field $\phi$, we simply need to quantise the simple harmonic oscillators. We do so in our non-standard framework. 

Consider the object $\SpaceH$ of $\starHilbCategory$ defined as follows, where $\tau$ is some infinite non-standard natural and $\ket{n}_{n \in \starNaturals}$ is the chosen orthonormal basis for $\nonstd{\Ltwo{\naturals}}$: 
\begin{equation}
	\SpaceH := \Big(\nonstd{\Ltwo{\naturals}}, \sum_{n=0}^{\tau} \ket{n}\bra{n}\Big)
\end{equation}
We will think of $\SpaceH$ as the non-standard counterpart for a quantum harmonic oscillator: the states $\ket{n}$ correspond to energy eigenstates for the oscillator, and we extended our range of energy values all the way up to some infinite natural $\tau$. We define the \textbf{ladder operators} $a$ and $a^\dagger$ on $\SpaceH$ as follows:
\begin{equation}
	a \ket{n} =
		\begin{cases}
			0 &\text{ if } n = 0 \\
			\sqrt{n} \ket{n-1} &\text{ otherwise}
		\end{cases}
	\hspace{3cm} 
	a^\dagger \ket{n} = 
		\begin{cases}
			0 &\text{ if } n = \tau \\
			\sqrt{n+1} \ket{n+1} &\text{ otherwise}
		\end{cases}
\end{equation}
It is easy to check that these operators satisfy the usual canonical commutation relations, up to a correction factor accounting for the truncation of energy above the infinite $\tau$:
\begin{equation}
	[a,a^\dagger] = \id{} - (\tau+1)\ket{\tau}\bra{\tau}
\end{equation}
When restricting ourselves to finite energy states, these operators are exactly the ladder operators for the quantum harmonic oscillator. In fact, the commutator above differs from the usual expression only up to a scalar multiple of the $\ket{\tau}\bra{\tau}$ rank-1 projector, which will be sent to a (scalar multiple of) itself by most operators of interest in this work. As a consequence, we define the following equivalence relation $=_{\tau}$ on operators $\SpaceH \rightarrow \SpaceH$, which we read ``equal up to $\tau$'':
\begin{equation}
	f =_\tau g \text{ if and only if } f-g \propto \ket{\tau}\bra{\tau} 
\end{equation}
In particular, the commutation relation for the ladder operators can be re-written as:
\begin{equation}
	[a,a^\dagger] =_\tau \id{}
\end{equation}
We will also extend the equivalence relation $=_\tau$ to maps which we have explicitly decomposed in the form $f,g:\SpaceH^{\otimes X} \otimes \SpaceG \rightarrow \SpaceH^{\otimes X} \otimes \SpaceK$ by setting:
\begin{equation}
	f =_\tau g \text{ if and only if } f-g = (\ket{\tau}\bra{\tau})^{\otimes X} \otimes h \text{ for some } h : \SpaceG \rightarrow \SpaceK
\end{equation}

We then proceed to define the \textbf{number operator} $N:=a^\dagger a$, and we obtain the usual property and commutators for it (exact equalities this time):
\begin{equation}
	N \ket{n} = n \ket{n} \hspace{3cm} [N,a^\dagger] = a^\dagger \hspace{3cm} [N,a] = -a
\end{equation}
The number operator is associated to a $\dagger$-SCFA on $\SpaceH$, the \textbf{number observable}, with $\ket{n}$ as classical states. For a quantum harmonic oscillator of energy $E_{\underline{p}}$, the Hamiltonian can finally be defined as:
\begin{equation}
	H := E_{\underline{p}} N
\end{equation}
Aside perhaps for the correction term in the canonical commutation relation, this is exactly what we would expect the non-standard version of the quantum harmonic oscillator to look like, and the traditional quantum harmonic oscillator is recovered by restricting to states of finite energy.

\subsection{Quantum fields}

We saw before that a solution to the Klein-Gordon can be interpreted to describe a field of simple harmonic oscillators at each point $\underline{p} \in \reals^3$ of momentum space, vibrating independently with energies given by the relativistic dispersion relation $E_{\underline{p}} = \sqrt{(|\underline{p}|c)^2 + (mc^2)^2}$. The natural quantisation of such a scenario involves considering independent quantum harmonic oscillators at each point $\underline{p} \in \reals^3$ of momentum space, i.e. an infinite direct product of separable Hilbert spaces over the 3-dimensional continuum. Because such a space would be mathematically unwieldy, and because only finite energy states are deemed to be physically interesting, the infinite direct product of quantum harmonic oscillators is never constructed, and the Fock space is considered instead. The Fock space is the Hilbert space of joint states for the quantum harmonic oscillators which is spanned by those separable states involving only finitely many oscillators not in their ground state: the state $\ket{n}$ for the oscillator at point $\underline{p} \in \reals^3$ is considered to count the number of quantum particles with definite momentum $\underline{p}$, and the Fock space is spanned by all states containing finitely many particles.

Within our non-standard framework, we don't have to worry about infinite tensor products, and we don't have to restrict ourselves to finite energy states or finite number of particles: as a consequence, we quantise the real scalar field $\phi$ by constructing the field of quantum harmonic oscillators in all its glory. This can be done by considering the space  $\SpaceH^{\otimes \frac{1}{\omega_{ir}}\starIntegersModPow{2\omega+1}{3}}$ defined above: we discretise momentum space to an infinite lattice $\frac{1}{\omega_{ir}}\starIntegersModPow{2\omega+1}{3}$ of infinitesimal mesh $1/\omega_{ir}$, and we place an independent quantum harmonic oscillator $\SpaceH$ at each point of the lattice (with varying frequency $\nu_{\underline{p}}$). 

For each $\underline{p} \in  \frac{1}{\omega_{ir}}\starIntegersModPow{2\omega+1}{3}$, we write $a_{\underline{p}}$ and $a^\dagger_{\underline{p}}$ for the ladder operators acting on the quantum Harmonic oscillator at $\underline{p}$ (tensored with the identity on all other oscillators), and $\ket{n @ \underline{p}}_{n=0}^\tau$ for the orthonormal basis of the oscillator at $\underline{p}$. We define the rescaled versions $a(\underline{p}):=\sqrt{\omega_{ir}^3}\;\;a_{\underline{p}}$ and $a^\dagger(\underline{p}):=\sqrt{\omega_{ir}^3}\;\;a^\dagger_{\underline{p}}$, which satisfy the following commutation relations:
\begin{align}
		[a(\underline{p}),a^\dagger(\underline{q})] &=_\tau \omega_{ir}^3 \delta_{\underline{p},\underline{q}}\id{} \nonumber\\
		[a(\underline{p}),a(\underline{q})] &= 0 \nonumber\\ 
		[a^\dagger(\underline{p}),a^\dagger(\underline{q})] &=0
\end{align}
where $\delta_{\underline{p},\underline{q}}$ is a symbol, defined to be equal to $1$ if $\underline{p}=\underline{q}$ and equal to $0$ otherwise.

The usual field operators $\phi(\underline{x})$ and $\pi(\underline{x})$ for a real scalar field can be defined from $a(\underline{p})$ and $a^\dagger(\underline{p})$ through the following discretised integral, for all points $\underline{x} \in \frac{1}{\omega_{uv}}\starIntegersModPow{2\omega+1}{3}$ in space:
	%\marginSG{Add that these are the operators for the real scalar field, and later on we will explain show (graphically) how to build the ones for the complex scalar field (remember to actually do this).}
	\footnote{Note that in going from momentum $\underline{p}$ to position $\underline{x}$ we have swapped the UV infinity $\omega_{ir}$ and the IR infinite $\omega_{uv}$, so that the lattice for position has infinitesimal mesh $\frac{1}{\omega_{uv}}$ instead of $\frac{1}{\omega_{ir}}$. See \cite{GogiosoGenovese2017} for further details about this effect.}
	\footnote{The factor $\frac{1}{\omega_{ir}}$ is the discretised non-standard counterpart of the differential $d\underline{p}$ in the standard formulation.}
\begin{align}
	\phi(\underline{x}) &:= \sum_{\underline{p}} \frac{1}{\omega_{ir}^3} \frac{1}{\sqrt{2E_{\underline{p}}}} \Big[ a(\underline{p})e^{i2\pi \underline{p}\cdot \underline{x}} + a^\dagger(\underline{p})e^{-i2\pi \underline{p}\cdot \underline{x}} \Big] \nonumber \\
	\pi(\underline{x}) &:= \sum_{\underline{p}} \frac{1}{\omega_{ir}^3} (-i)\frac{\sqrt{E_{\underline{p}}}}{2} \Big[ a(\underline{p})e^{i2\pi \underline{p}\cdot \underline{x}} - a^\dagger(\underline{p})e^{-i2\pi \underline{p}\cdot \underline{x}} \Big] 
\end{align}
The field operators satisfy commutation relations similar to the ones of the rescaled ladder operators:
\begin{align}
		[\phi(\underline{x}),\pi(\underline{y})] &=_\tau \omega_{uv}^3 \delta_{\underline{x},\underline{y}}\id{} \nonumber\\ % TODO: check this 
		[\phi(\underline{x}),\phi(\underline{y})] &= 0 \nonumber\\ 
		[\pi(\underline{x}),\pi(\underline{y})] &=0
\end{align}
Operators for the complex fields can be defined similarly, using a pair of harmonic oscillators (one for particles, one for anti-particles).

For every $\textbf{n}:\frac{1}{\omega_{ir}}\starIntegersModPow{2\omega+1}{3} \rightarrow \{0,...,\tau\}$, we can define the state $\ket{\textbf{n}} := \otimes_{\underline{p}} \ket{\textbf{n}(\underline{p}) @\underline{p}}$. In particular, the state $\ket{\textbf{0}}$ is the vacuum, and the single-particle states can be defined as usual by:
\begin{equation}
	\ket{\underline{p}} := a^\dagger(\underline{p}) \ket{\textbf{0}} 	
\end{equation} 
The discretised integral of the (rescaled) number observables for all quantum harmonic oscillators at all points $\underline{p} \in \frac{1}{\omega_{ir}}\starIntegersModPow{2\omega+1}{3}$ of momentum space gives rise to the \textbf{number operator} $N$ on $\SpaceH^{\otimes \frac{1}{\omega_{ir}}\starIntegersModPow{2\omega+1}{3}}$:
\begin{equation}
	N  := \sum_{\underline{p}} \frac{1}{\omega_{ir}^3} a^\dagger({\underline{p}}) a({\underline{p}}) = \sum_{\underline{p}} a^\dagger_{\underline{p}}a_{\underline{p}} =\sum_{\textbf{n}}  \Big(\sum_{\underline{p}}\textbf{n}(\underline{p})\Big) \ket{\textbf{n}}\bra{\textbf{n}}
\end{equation}
The \textbf{Hamiltonian} for the quantum field is similarly obtained as a discretised integral:
\begin{equation}
	H :=  \sum_{\underline{p}} \frac{1}{\omega_{ir}^3} E_{\underline{p}} a^\dagger({\underline{p}}) a({\underline{p}}) = \sum_{\underline{p}} E_{\underline{p}} a^\dagger_{\underline{p}} a_{\underline{p}} =\sum_{\textbf{n}}  \Big(\sum_{\underline{p}} E_{\underline{p}}\textbf{n}(\underline{p})\Big) \ket{\textbf{n}}\bra{\textbf{n}}
\end{equation}
The traditional Fock space is recovered by considering the states $\ket{\textbf{n}}$ with finite energy $\bra{\textbf{n}} H \ket{\textbf{n}}$ (i.e. those with a finite number of particles, all having finite momenta). The corresponding number of particles at a standard point $\underline{q} \in \reals^3$ of standard momentum space, which we will denote by $\stdpart{\textbf{n}}(\underline{q})$, is then given by the following expression:
\begin{equation}
	\stdpart{\textbf{n}}(\underline{q}) := \hspace{-3mm} \sum_{\substack{\underline{p} \in  \frac{1}{\omega_{ir}}\starIntegersModPow{2\omega+1}{3} \\\text{such that } \stdpart{\underline{p}}= \underline{q}}} \hspace{-3mm} \textbf{n}(\underline{p})
\end{equation}

\subsection{Relativistic quantum fields}

Because we are only interested in on-shell physical states satisfying the relativistic dispersion relation, we have defined our fields directly on 3-momentum values $\underline{p} \in \frac{1}{\omega_{ir}}\starIntegersModPow{2\omega+1}{3}$, since the $p_0 = E_{\underline{p}}$ values can be uniquely determined from them. Unfortunately, however, the volume element $d^3\!\underline{p} = \frac{1}{\omega_{ir}^3}$ which we have used in our integrals is not Lorentz-invariant, so we need to fix it. The Lorentz-invariant volume element for 3-momenta is $d^3\!\underline{p}\;\frac{1}{2E_{\underline{p}}} = \frac{1}{\omega_{ir}^3}\frac{1}{2E_{\underline{p}}}$
	\footnote{
	To see this, note that the 4-momentum volume element $d^4\!p$ is Lorentz-invariant, and that so is the on-shell 4-momentum volume element $d^4\!p \; \delta(p_0-E_{\underline{p}}) |_{p_0 >0}$. A quick differentiation yields:
	\[
		d^4\!p \;\delta(p_0-E_{\underline{p}}) |_{p_0 >0} = d^3\!\underline{p}\;\; \frac{1}{2p_0}|_{p_0 = E_p}
	\].
	} 
and we re-normalise our states and operators to make the extra factor $\frac{1}{2E_{\underline{p}}}$ apparent in our integrals. We define the \textbf{relativistically normalised} ladder operators and one-particle states by:
\begin{equation}
	a(p) := \sqrt{2E_{\underline{p}}} \;\;a(\underline{p})
	\hspace{2cm}
	a^\dagger(p) := \sqrt{2E_{\underline{p}}}\;\; a^\dagger(\underline{p})
	\hspace{2cm}
	\ket{p} := a^\dagger(p) \ket{\textbf{0}} = \sqrt{2E_{\underline{p}}}\ket{\underline{p}}
\end{equation}
This means that the identity on one-particle states is the self-evidently Lorentz-invariant discretised integral $\sum_{\underline{p}} \frac{1}{\omega_{ir}^3}\frac{1}{2E_{\underline{p}}} \ket{p}\bra{p}$, and that the field operator $\phi(\underline{x})$ can be re-written in terms of Lorentz-invariant volume element and Lorentz-invariant ladder operators as follows:
\begin{equation}
	\phi(\underline{x}) = \sum_{\underline{p}} \frac{1}{\omega_{ir}^3} \frac{1}{2E_{\underline{p}}} \Big[ a(p)e^{i2\pi \underline{p}\cdot \underline{x}} + a^\dagger({p})e^{-i2\pi \underline{p}\cdot \underline{x}} \Big] 
\end{equation}
Unfortunately, the expression for the field operator is not yet Lorentz-invariant, because of the imaginary exponential factors. This will soon be fixed when we start working in the Heisenberg picture.

\subsection{Coherently-controlled creation/destruction operators}

Before moving on to the Heisenberg picture, we take a brief detour to talk about position-momentum duality in our setting. We move away from the traditional approach to ladder/field operators---seen as families of operators classically parametrised by position/momentum values---and we switch to the coherent approach of \cite{CategoricalQuantumDynamics}, working within the object $\starComplexs[\frac{1}{\omega_{ir}} \starIntegersModPow{2\omega+1}{3}]$ of $\starHilbCategory$. From a process-theoretic perspective, working with the parameters $\underline{x}$ and $\underline{p}$ corresponds to thinking of processes which are classically controlled by points in the position and momentum spaces (or, more generally, by probability distributions over them); working in $\starComplexs[\frac{1}{\omega_{ir}} \starIntegersModPow{2\omega+1}{3}]$, on the other hand, corresponds to thinking of processes which are `coherently' controlled by wavefunctions over the position and momentum spaces. Because of (non-standard) Pontryagin duality, this has the advantage that the same parameter space can be used for both position and momentum, allowing a number of useful duality results to be easily formulated.

Recall that in $\starComplexs[\frac{1}{\omega_{ir}} \starIntegersModPow{2\omega+1}{3}]$ we have an orthogonal basis of momentum eigenstates $(\ket{\goodchi_{\underline{p}}})_{\underline{p} \in \frac{1}{\omega_{ir}} \starIntegersModPow{2\omega+1}{3}}$, and an associated quasi-special commutative dagger Frobenius algebra. In the context of quantum field theory, we need to take relativistic normalisation into account, so we consider the orthogonal basis of \textbf{relativistically normalised momentum eigenstates} instead:
\begin{equation}
	\ket{\goodchi_p} := \sqrt{2E_{\underline{p}}} \ket{\goodchi_{\underline{p}}}
\end{equation}
This is associated to a commutative dagger Frobenius algebra $\hbox{\begin{tikzpicture} [scale=1.2,transform shape] %% DO NOT CHANGE

\def\deltax{0.3} %% CAN BE CHANGED
\def\deltay{0.5} %% DO NOT CHANGE

%\path[use as bounding box] (-\deltax,-\deltay) rectangle (\deltax,\deltay);

\node [dot, fill=\Zbwcolour] (mult) at (0,0) {};
%\node (mult_label_out) at (0,+\deltay) {};
%\draw[-] (mult) to (mult_label_out);

%\draw (current bounding box.south west) rectangle (current bounding box.north east);
\end{tikzpicture}
}\!\!$, which is no longer quasi-special (because the states have different square norms \cite{CoeckePavlovicVicary2013}). To take relativistic normalisation into account, we re-define the orthogonal basis of position eigenstates $(\ket{\delta_{\underline{x}}})_{\underline{x} \in \frac{1}{\omega_{uv}} \starIntegersModPow{2\omega+1}{3}}$ using the relativistically normalised momentum eigenstates in place of the original ones:
\begin{equation}
	\ket{\delta_{\underline{x}}} 
	:= 
	\sum_{\underline{p} \in  \frac{1}{\omega_{ir}}\starIntegersModPow{2\omega+1}{n}} 
	\!\!\!\!\frac{1}{\omega_{ir}^3}\frac{1}{2E_{\underline{p}}}\;\;
	e^{-i 2\pi \, \underline{p}\cdot \underline{x}} \ket{\goodchi_{p}}
\end{equation}
The position eigenstates have an associated commutative dagger Frobenius algebra $\hbox{\begin{tikzpicture} [scale=1.2,transform shape] %% DO NOT CHANGE

\def\deltax{0.3} %% CAN BE CHANGED
\def\deltay{0.5} %% DO NOT CHANGE

%\path[use as bounding box] (-\deltax,-\deltay) rectangle (\deltax,\deltay);

\node [dot, fill=\Dcolour] (mult) at (0,0) {};
%\node (mult_label_out) at (0,+\deltay) {};
%\draw[-] (mult) to (mult_label_out);

%\draw (current bounding box.south west) rectangle (current bounding box.north east);
\end{tikzpicture}
}\!\!$, and the pair $(\hbox{\begin{tikzpicture} [scale=1.2,transform shape] %% DO NOT CHANGE

\def\deltax{0.3} %% CAN BE CHANGED
\def\deltay{0.5} %% DO NOT CHANGE

%\path[use as bounding box] (-\deltax,-\deltay) rectangle (\deltax,\deltay);

\node [dot, fill=\Dcolour] (mult) at (0,0) {};
%\node (mult_label_out) at (0,+\deltay) {};
%\draw[-] (mult) to (mult_label_out);

%\draw (current bounding box.south west) rectangle (current bounding box.north east);
\end{tikzpicture}
}\!\!,\hbox{\begin{tikzpicture} [scale=1.2,transform shape] %% DO NOT CHANGE

\def\deltax{0.3} %% CAN BE CHANGED
\def\deltay{0.5} %% DO NOT CHANGE

%\path[use as bounding box] (-\deltax,-\deltay) rectangle (\deltax,\deltay);

\node [dot, fill=\Zbwcolour] (mult) at (0,0) {};
%\node (mult_label_out) at (0,+\deltay) {};
%\draw[-] (mult) to (mult_label_out);

%\draw (current bounding box.south west) rectangle (current bounding box.north east);
\end{tikzpicture}
}\!\!)$ is strongly complementary (capturing features of position-momentum duality such as the Weyl CCRs \cite{CategoricalQuantumDynamics}).

Instead of working with the $\underline{p}$-parametrised ladder operators and the $\underline{x}$-parametrised field operator, we define the following morphism $\gamma^\dagger:\SpaceH^{\otimes\frac{1}{\omega_{ir}}\starIntegersModPow{2\omega+1}{3}} \otimes \starComplexs[\frac{1}{\omega_{ir}} \starIntegersModPow{2\omega+1}{3}] \longrightarrow \SpaceH^{\otimes\frac{1}{\omega_{ir}}\starIntegersModPow{2\omega+1}{3}}$, which we refer to as the \textbf{coherently controlled creation operator}:
\begin{equation} 
	\begin{tikzpicture}
	\begin{pgfonlayer}{nodelayer}
		\node [style=box,minimum height = 12mm, minimum width = 8mm] (box) at (0, 0) {$\gamma^\dagger$};
		\node [style=none] (inl) at (-2,0.6) {};
		\node [style=none] (box_inl) at (0,0.6) {};
		\node [style=none] (inr) at (-2,-0.6) {};
		\node [style=none] (box_inr) at (0,-0.6) {};
		\node [style=none] (out) at (2,0) {};
		\node [style=none] (box_out) at (0,0) {};
	\end{pgfonlayer}
	\begin{pgfonlayer}{edgelayer}
		\draw[-] (inl) to (box_inl); 
		\draw[-,draw=red] (inr) to (box_inr);
		\draw[-] (box_out) to (out);
	\end{pgfonlayer}
\end{tikzpicture}
	\hspace{3mm}
	:=
	\hspace{3mm}
	\sum_{\underline{p}}
	\frac{1}{\omega_{ir}^3}\frac{1}{2E_{\underline{p}}} 
	\begin{tikzpicture}
	\begin{pgfonlayer}{nodelayer}
		\node [style=box] (box) at (1, 0.6) {$a^\dagger({p})$};
		\node [style=none] (in) at (-2,0.6) {};
		\node [style=none] (out) at (4,0.6) {};
		\node [style=effect,draw=red,text=red] (effect) at (-0.5,-0.6) {$\goodchi_{p}$};
		\node [style=none] (effect_in) at (-2,-0.6) {};
	\end{pgfonlayer}
	\begin{pgfonlayer}{edgelayer}
		\draw[-] (in) to (box);
		\draw[-,out=0,in=180] (box) to (out);
		\draw[-,draw=red] (effect_in) to (effect);
	\end{pgfonlayer}
\end{tikzpicture}
\end{equation}
The (relativistically normalised) ladder operators can be recovered by applying the coherently controlled creation/destruction operators $\gamma^\dagger$ and $\gamma$ to the (relativistically normalised) momentum eigenstates\footnote{We used light gray lines/borders in place of black lines/borders to distinguish between the field space $\SpaceH^{\otimes\frac{1}{\omega_{ir}}\starIntegersModPow{2\omega+1}{3}}$ and the controlling parameter space $\starComplexs[\frac{1}{\omega_{ir}} \starIntegersModPow{2\omega+1}{3}]$.}:
\begin{align}
	\begin{tikzpicture}
	\begin{pgfonlayer}{nodelayer}
		\node [style=box,minimum height = 12mm, minimum width = 8mm] (box) at (0, 0) {$\gamma$};
		\node [style=none] (in) at (-2.5,0) {};
		\node [style=none] (box_in) at (0,0) {};
		\node [style=state,draw=red,text=red] (inr) at (-2,-1.6) {$\goodchi_{{p}}$};
		\node [style=none] (dot_inl) at (1,-0.6) {};
		\node [style=none] (dot_inr) at (1,-1.6) {};
		\node [style=Zbwdot,draw=red] (dot) at (1.5,-1.1) {};
		\node [style=none] (box_outr) at (0,-0.6) {};
		\node [style=none] (outl) at (2,0.6) {};
		\node [style=none] (box_outl) at (0,0.6) {};
	\end{pgfonlayer}
	\begin{pgfonlayer}{edgelayer}
		\draw[-] (in) to (box_in); 
		\draw[-,draw=red] (inr) to (dot_inr.center);
		\draw[-,draw=red] (box_outr) to (dot_inl.center);
		\draw[-,out=0,in=90,draw=red] (dot_inl.center) to (dot);
		\draw[-,out=0,in=-90,draw=red] (dot_inr.center) to (dot);
		\draw[-] (box_outl) to (outl);
	\end{pgfonlayer}
\end{tikzpicture}
	=
	\begin{tikzpicture}
	\begin{pgfonlayer}{nodelayer}
		\node [style=box] (box) at (0, 0) {$a({p})$};
		\node [style=none] (in) at (-2,0) {};
		\node [style=none] (box_in) at (0,0) {};
		\node [style=none] (out) at (2,0) {};
		\node [style=none] (box_out) at (0,0) {};
	\end{pgfonlayer}
	\begin{pgfonlayer}{edgelayer}
		\draw[-] (in) to (box_in);
		\draw[-] (box_out) to (out);
	\end{pgfonlayer}
\end{tikzpicture}
	&\hspace{2cm}
	\begin{tikzpicture}
	\begin{pgfonlayer}{nodelayer}
		\node [style=box,minimum height = 12mm, minimum width = 8mm] (box) at (0, 0) {$\gamma$};
		\node [style=none] (in) at (-2.5,0) {};
		\node [style=none] (box_in) at (0,0) {};
		\node [style=state,draw=red,text=red] (inr) at (-2,-1.6) {$\goodchi_{\underline{p}}$};
		\node [style=none] (dot_inl) at (1,-0.6) {};
		\node [style=none] (dot_inr) at (1,-1.6) {};
		\node [style=Zbwdot,draw=red] (dot) at (1.5,-1.1) {};
		\node [style=none] (box_outr) at (0,-0.6) {};
		\node [style=none] (outl) at (2,0.6) {};
		\node [style=none] (box_outl) at (0,0.6) {};
	\end{pgfonlayer}
	\begin{pgfonlayer}{edgelayer}
		\draw[-] (in) to (box_in); 
		\draw[-,draw=red] (inr) to (dot_inr.center);
		\draw[-,draw=red] (box_outr) to (dot_inl.center);
		\draw[-,out=0,in=90,draw=red] (dot_inl.center) to (dot);
		\draw[-,out=0,in=-90,draw=red] (dot_inr.center) to (dot);
		\draw[-] (box_outl) to (outl);
	\end{pgfonlayer}
\end{tikzpicture}
	=
	\begin{tikzpicture}
	\begin{pgfonlayer}{nodelayer}
		\node [style=box] (box) at (0, 0) {$a(\underline{p})$};
		\node [style=none] (in) at (-2,0) {};
		\node [style=none] (box_in) at (0,0) {};
		\node [style=none] (out) at (2,0) {};
		\node [style=none] (box_out) at (0,0) {};
	\end{pgfonlayer}
	\begin{pgfonlayer}{edgelayer}
		\draw[-] (in) to (box_in);
		\draw[-] (box_out) to (out);
	\end{pgfonlayer}
\end{tikzpicture}
	\nonumber \\
	\begin{tikzpicture}
	\begin{pgfonlayer}{nodelayer}
		\node [style=box,minimum height = 12mm, minimum width = 8mm] (box) at (0, 0) {$\gamma^\dagger$};
		\node [style=none] (inl) at (-2.5,0.6) {};
		\node [style=none] (box_inl) at (0,0.6) {};
		\node [style=state,draw=red,text=red] (inr) at (-2,-0.6) {$\goodchi_{p}$};
		\node [style=none] (box_inr) at (0,-0.6) {};
		\node [style=none] (out) at (2,0) {};
		\node [style=none] (box_out) at (0,0) {};
	\end{pgfonlayer}
	\begin{pgfonlayer}{edgelayer}
		\draw[-] (inl) to (box_inl); 
		\draw[-,draw=red] (inr) to (box_inr);
		\draw[-] (box_out) to (out);
	\end{pgfonlayer}
\end{tikzpicture}
	=
	\begin{tikzpicture}
	\begin{pgfonlayer}{nodelayer}
		\node [style=box] (box) at (0, 0) {$a^\dagger({p})$};
		\node [style=none] (in) at (-2,0) {};
		\node [style=none] (box_in) at (0,0) {};
		\node [style=none] (out) at (2,0) {};
		\node [style=none] (box_out) at (0,0) {};
	\end{pgfonlayer}
	\begin{pgfonlayer}{edgelayer}
		\draw[-] (in) to (box_in);
		\draw[-] (box_out) to (out);
	\end{pgfonlayer}
\end{tikzpicture}
	&\hspace{2cm}
	\begin{tikzpicture}
	\begin{pgfonlayer}{nodelayer}
		\node [style=box,minimum height = 12mm, minimum width = 8mm] (box) at (0, 0) {$\gamma^\dagger$};
		\node [style=none] (inl) at (-2.5,0.6) {};
		\node [style=none] (box_inl) at (0,0.6) {};
		\node [style=state,draw=red,text=red] (inr) at (-2,-0.6) {$\goodchi_{\underline{p}}$};
		\node [style=none] (box_inr) at (0,-0.6) {};
		\node [style=none] (out) at (2,0) {};
		\node [style=none] (box_out) at (0,0) {};
	\end{pgfonlayer}
	\begin{pgfonlayer}{edgelayer}
		\draw[-] (inl) to (box_inl); 
		\draw[-,draw=red] (inr) to (box_inr);
		\draw[-] (box_out) to (out);
	\end{pgfonlayer}
\end{tikzpicture}
	=
	\begin{tikzpicture}
	\begin{pgfonlayer}{nodelayer}
		\node [style=box] (box) at (0, 0) {$a^\dagger(\underline{p})$};
		\node [style=none] (in) at (-2,0) {};
		\node [style=none] (box_in) at (0,0) {};
		\node [style=none] (out) at (2,0) {};
		\node [style=none] (box_out) at (0,0) {};
	\end{pgfonlayer}
	\begin{pgfonlayer}{edgelayer}
		\draw[-] (in) to (box_in);
		\draw[-] (box_out) to (out);
	\end{pgfonlayer}
\end{tikzpicture}
\end{align}
The commutation relations for the ladder operators can be written graphically in terms of $\gamma^\dagger$ and $\gamma$:
\begin{equation}\label{gammaMapCommutator}
	\begin{tikzpicture}
	\begin{pgfonlayer}{nodelayer}
		\node [style=box,minimum height = 12mm, minimum width = 8mm] (gamma) at (-2, 0) {$\gamma^\dagger$};
		\node [style=box,minimum height = 12mm, minimum width = 8mm] (gammad) at (+2,0) {$\gamma$};

		\node [style=none] (inl) at (-4.5,0.6) {};
		\node [style=none] (inr) at (-4.5,-0.6) {};		
		\node [style=none] (inrr) at (-4.5,-1.6) {};
		\node [style=none] (gammainl) at (-2,0.6) {};
		\node [style=none] (gammainr) at (-2,-0.6) {};

		\node [style=none] (outl) at (4.5,0.6) {};
		\node [style=none] (dotinl) at (3,-0.6) {};
		\node [style=none] (gammadoutl) at (+2,0.6) {};
		\node [style=none] (gammadoutr) at (+2,-0.6) {};
		\node [style=none] (dotinr) at (3,-1.6) {};
		\node [style=dot,draw=red] (dot) at (3.5,-1.1){};
	\end{pgfonlayer}
	\begin{pgfonlayer}{edgelayer}
		\draw[-] (gamma) to (gammad); 
		\draw[-] (inl) to (gammainl); 
		\draw[-,draw=red] (inr) to (gammainr); 
		\draw[-] (gammadoutl) to (outl); 
		\draw[-,draw=red] (gammadoutr) to (dotinl.center); 
		\draw[-,out=0,in=90,draw=red] (dotinl.center) to (dot);
		\draw[-,out=0,in=-90,draw=red] (dotinr.center) to (dot);
		\draw[-,draw=red] (inrr) to (dotinr.center);
	\end{pgfonlayer}
\end{tikzpicture}
	\hspace{3mm}
	-
	\hspace{3mm}
	\begin{tikzpicture}
	\begin{pgfonlayer}{nodelayer}
		\node [style=box,minimum height = 12mm, minimum width = 8mm] (gamma) at (-2, 0) {$\gamma$};
		\node [style=box,minimum height = 12mm, minimum width = 8mm] (gammad) at (+2,0) {$\gamma^\dagger$};
		\node [style=none] (gammal) at (-2,0.6) {};
		\node [style=none] (gammadl) at (2,0.6) {};

		\node [style=none] (inl) at (-4.5,0) {};
		\node [style=none] (inr) at (-4.5,-0.6) {};	
		\node [style=none] (inr2) at (-4,-0.6) {};	
		\node [style=none] (inr3) at (-2,-2.1) {};	
		\node [style=none] (inr4) at (0,-2.1) {};		
		\node [style=none] (inr5) at (1,-0.6) {};	
		\node [style=none] (inrr) at (-4.5,-1.6) {};
		\node [style=none] (gammainl) at (-2,0) {};
		\node [style=none] (gammainr) at (2,-0.6) {};

		\node [style=none] (outl) at (4.5,0) {};
		\node [style=none] (dotinl) at (-1,-0.6) {};
		\node [style=none] (gammadoutl) at (+2,0) {};
		\node [style=none] (gammadoutr) at (-2,-0.6) {};
		\node [style=none] (dotinr) at (-1,-1.6) {};
		\node [style=dot,draw=red] (dot) at (-0.5,-1.1){};
	\end{pgfonlayer}
	\begin{pgfonlayer}{edgelayer}
		\draw[-] (gammal) to (gammadl); 
		\draw[-] (inl) to (gammainl); 
		\draw[-,draw=red] (inr) to (inr2.center); 
		\draw[-,draw=red,out=0,in=180] (inr2.center) to (inr3.center); 
		\draw[-,draw=red,out=0,in=180] (inr3.center) to (inr4.center); 
		\draw[-,draw=red,out=0,in=180] (inr4.center) to (inr5.center); 
		\draw[-,draw=red,out=0,in=180] (inr5.center) to (gammainr); 
		% \draw[-,draw=red] (inr) to (gammainr); 
		\draw[-] (gammadoutl) to (outl); 
		\draw[-,draw=red] (gammadoutr) to (dotinl.center); 
		\draw[-,out=0,in=90,draw=red] (dotinl.center) to (dot);
		\draw[-,out=0,in=-90,draw=red] (dotinr.center) to (dot);
		\draw[-,draw=red] (inrr) to (dotinr.center);
	\end{pgfonlayer}
\end{tikzpicture}
	\hspace{4mm}
	=_\tau
	\hspace{4mm}
	\begin{tikzpicture}
	\begin{pgfonlayer}{nodelayer}
		\node [style=none] (inl) at (-1,0.4) {};
		\node [style=none] (inr) at (-1,-0.6) {};		
		\node [style=none] (inrr) at (-1,-1.6) {};

		\node [style=none] (outl) at (3,0.4) {};
		\node [style=none] (dotinl) at (0,-0.6) {};
		\node [style=none] (dotinr) at (0,-1.6) {};
		\node [style=dot,draw=red] (dot) at (0.5,-1.1){};
	\end{pgfonlayer}
	\begin{pgfonlayer}{edgelayer}
		\draw[-] (inl) to (outl); 
		\draw[-,draw=red] (inr) to (dotinl.center); 
		\draw[-,out=0,in=90,draw=red] (dotinl.center) to (dot);
		\draw[-,out=0,in=-90,draw=red] (dotinr.center) to (dot);
		\draw[-,draw=red] (inrr) to (dotinr.center);
	\end{pgfonlayer}
\end{tikzpicture}
\end{equation}

We can use the additional degrees of freedom granted by the coherent approach to obtain the positive and negative frequency parts $\phi^+(\underline{x}),\phi^-(\underline{x}),$ of the field operator $\phi(\underline{x})$, by evaluating the $\gamma$ and $\gamma^\dagger$ map on the position eigenstates $\ket{\delta_{\underline{x}}}$ instead of the momentum eigenstates $\ket{\goodchi_{\underline{p}}}$:
\begin{align}
	\begin{tikzpicture}
	\begin{pgfonlayer}{nodelayer}
		\node [style=box,minimum height = 12mm, minimum width = 8mm] (box) at (0, 0) {$\gamma$};
		\node [style=none] (in) at (-2.5,0) {};
		\node [style=none] (box_in) at (0,0) {};
		\node [style=state,draw=red,text=red] (inr) at (-2,-1.6) {$\delta_{\underline{x}}$};
		\node [style=none] (dot_inl) at (1,-0.6) {};
		\node [style=none] (dot_inr) at (1,-1.6) {};
		\node [style=Xbwdot,draw=red] (dot) at (1.5,-1.1) {};
		\node [style=none] (box_outr) at (0,-0.6) {};
		\node [style=none] (outl) at (2,0.6) {};
		\node [style=none] (box_outl) at (0,0.6) {};
	\end{pgfonlayer}
	\begin{pgfonlayer}{edgelayer}
		\draw[-] (in) to (box_in); 
		\draw[-,draw=red] (inr) to (dot_inr.center);
		\draw[-,draw=red] (box_outr) to (dot_inl.center);
		\draw[-,out=0,in=90,draw=red] (dot_inl.center) to (dot);
		\draw[-,out=0,in=-90,draw=red] (dot_inr.center) to (dot);
		\draw[-] (box_outl) to (outl);
	\end{pgfonlayer}
\end{tikzpicture}
	&=
	\phi^{+}(\underline{x})
	:=
	\sum_{\underline{p}} \frac{1}{\omega_{ir}^3} \frac{1}{2E_{\underline{p}}} 
	a(p)e^{i2\pi \underline{p}\cdot \underline{x}} 
	\nonumber \\
	\begin{tikzpicture}
	\begin{pgfonlayer}{nodelayer}
		\node [style=box,minimum height = 12mm, minimum width = 8mm] (box) at (0, 0) {$\gamma^\dagger$};
		\node [style=none] (inl) at (-2.5,0.6) {};
		\node [style=none] (box_inl) at (0,0.6) {};
		\node [style=state,draw=red,text=red] (inr) at (-2,-0.6) {$\delta_{\underline{x}}$};
		\node [style=none] (box_inr) at (0,-0.6) {};
		\node [style=none] (out) at (2,0) {};
		\node [style=none] (box_out) at (0,0) {};
	\end{pgfonlayer}
	\begin{pgfonlayer}{edgelayer}
		\draw[-] (inl) to (box_inl); 
		\draw[-,draw=red] (inr) to (box_inr);
		\draw[-] (box_out) to (out);
	\end{pgfonlayer}
\end{tikzpicture}
	&=
	\phi^{-}(\underline{x})
	:=
	\sum_{\underline{p}} \frac{1}{\omega_{ir}^3} \frac{1}{2E_{\underline{p}}}  
	a^\dagger({p})e^{-i2\pi \underline{p}\cdot \underline{x}} 
	\nonumber \\
\end{align}
Finally, we can use the coherently-controlled operators to write the number operator diagrammatically as follows:
\begin{equation}
	\begin{tikzpicture}
	\begin{pgfonlayer}{nodelayer}
		\node [style=box,minimum height = 8mm, minimum width = 8mm] (gamma) at (0, 0) {N};
		\node [style=none] (in) at (-2.5,0) {};
		\node [style=none] (out) at (2.5,0) {};
	\end{pgfonlayer}
	\begin{pgfonlayer}{edgelayer}
	\draw[-] (in) to (out);
	\end{pgfonlayer}
\end{tikzpicture}
	\hspace{2mm}
	=
	\hspace{2mm}
	\begin{tikzpicture}
	\begin{pgfonlayer}{nodelayer}
		\node [style=box,minimum height = 12mm, minimum width = 8mm] (gamma) at (-2, 0) {$\gamma^\dagger$};
		\node [style=box,minimum height = 12mm, minimum width = 8mm] (gammad) at (+2,0) {$\gamma$};
		\node [style=none] (inl) at (-4.5,0.6) {};
		\node [style=none] (inr) at (-4.5,-0.6) {};		
		\node [style=none] (inrr) at (-4.5,-1.6) {};
		\node [style=none] (gammainl) at (-2,0.6) {};
		\node [style=none] (gammainr) at (-2,-0.6) {};

		\node [style=none] (outl) at (4.5,0.6) {};
		\node [style=none] (dotinl) at (3,-0.6) {};
		\node [style=none] (gammadoutl) at (+2,0.6) {};
		\node [style=none] (gammadoutr) at (+2,-0.6) {};
		\node [style=none] (dotinr) at (3,-1.6) {};
		\node [style=dot,draw=red] (dot) at (3.5,-1.1){};
		\node [style=dot,draw=red] (dotlow) at (-3.5,-1.1){};
		\node [style=none] (dotlowinr) at (-3,-1.6) {};
	\end{pgfonlayer}
	\begin{pgfonlayer}{edgelayer}
		\draw[-] (gamma) to (gammad); 
		\draw[-] (inl) to (gammainl); 
		% \draw[-,draw=red] (inr) to (gammainr); 
		\draw[-] (gammadoutl) to (outl); 
		\draw[-,draw=red] (gammadoutr) to (dotinl.center); 
		\draw[-,out=0,in=90,draw=red] (dotinl.center) to (dot);
		\draw[-,out=0,in=-90,draw=red] (dotinr.center) to (dot);
		\draw[-,out=180,in=90,draw=red] (gammainr) to (dotlow);
		\draw[-,out=180,in=-90,draw=red] (dotlowinr.center) to (dotlow);
		\draw[-,draw=red] (dotlowinr.center) to (dotinr.center);
	\end{pgfonlayer}
\end{tikzpicture}
\end{equation}

\subsection{The Heisenberg picture}

So far, we have not taken time into consideration. Just as 3-dimensional space was discretised by the 3-dimensional non-standard lattice $\frac{1}{\omega_{uv}} \starIntegersModPow{2\omega+1}{3}$, so we take time to be discretised by the 1-dimensional non-standard lattice $\frac{1}{\omega_{uv}} \starIntegersMod{2\omega+1}$. The action of time-translation on relativistically normalised momentum eigenstates $\ket{\goodchi_{{p}}} \in \starComplexs[\frac{1}{\omega_{ir}} \starIntegersModPow{2\omega+1}{3}]$ should be captured by the following unitary representation $\big(U_t\big)_{t \in \frac{1}{\omega_{uv}} \starIntegersMod{2\omega+1}}$:
\begin{equation}
	U_t := \sum_{\underline{p}} \frac{1}{\omega_{ir}^3}\frac{1}{2E_{\underline{p}}} \, e^{i2\pi E_{\underline{p}} t} \ket{\goodchi_{p}}\bra{\goodchi_{p}}
\end{equation}
Unfortunately, there is a snag: the one above is not a well-defined unitary representation! This is because, in general, the energy $E_{\underline{p}} = \sqrt{(|\underline{p}|c)^2+(mc^2)^2}$ defined by the relativistic dispersion relation will not take values $E_{\underline{p}} \in \frac{1}{\omega_{ir}} \starIntegersMod{2\omega+1}$ in the Pontryagin dual of the discretised time-translation group. We ensure that the energy takes suitable values by applying an infinitesimal correction. Specifically, we take natural units in which $c=1$, we assume that $m \in \frac{1}{\omega_{ir}} \starIntegersModPow{2\omega+1}{3}$ and we redefine $E_{\underline{p}}$ to be:
\begin{equation}
	E_{\underline{p}} := \frac{1}{\omega_{ir}} \Big\lfloor\omega_{ir}\sqrt{|\underline{p}|^2+m^2}\Big\rfloor
\end{equation}
This makes the unitary representation $\big(U_t\big)_{t \in \frac{1}{\omega_{uv}} \starIntegersMod{2\omega+1}}$ above well-defined, at the cost of introducing some energy level degeneracy between some infinitesimally close momentum values.
	\footnote{The exact infinitesimal extent of this degeneracy depends, rather interestingly, on both $m$ and the ratio $\omega_{uv}/\omega_{ir}$.}

Just as positions and momenta were associated to a strongly complementary pair $(\hbox{\begin{tikzpicture} [scale=1.2,transform shape] %% DO NOT CHANGE

\def\deltax{0.3} %% CAN BE CHANGED
\def\deltay{0.5} %% DO NOT CHANGE

%\path[use as bounding box] (-\deltax,-\deltay) rectangle (\deltax,\deltay);

\node [dot, fill=\Dcolour] (mult) at (0,0) {};
%\node (mult_label_out) at (0,+\deltay) {};
%\draw[-] (mult) to (mult_label_out);

%\draw (current bounding box.south west) rectangle (current bounding box.north east);
\end{tikzpicture}
}\!\!,\hbox{\begin{tikzpicture} [scale=1.2,transform shape] %% DO NOT CHANGE

\def\deltax{0.3} %% CAN BE CHANGED
\def\deltay{0.5} %% DO NOT CHANGE

%\path[use as bounding box] (-\deltax,-\deltay) rectangle (\deltax,\deltay);

\node [dot, fill=\Zbwcolour] (mult) at (0,0) {};
%\node (mult_label_out) at (0,+\deltay) {};
%\draw[-] (mult) to (mult_label_out);

%\draw (current bounding box.south west) rectangle (current bounding box.north east);
\end{tikzpicture}
}\!\!)$ of quasi-special commutative dagger Frobenius algebras on $\starComplexs[\frac{1}{\omega_{ir}} \starIntegersModPow{2\omega+1}{3}]$, so time and energy are associated to a strongly complementary pair $(\hbox{\begin{tikzpicture} [scale=1.2,transform shape] %% DO NOT CHANGE

\def\deltax{0.3} %% CAN BE CHANGED
\def\deltay{0.5} %% DO NOT CHANGE

%\path[use as bounding box] (-\deltax,-\deltay) rectangle (\deltax,\deltay);

\node [dot, fill=\Dcolour] (mult) at (0,0) {};
%\node (mult_label_out) at (0,+\deltay) {};
%\draw[-] (mult) to (mult_label_out);

%\draw (current bounding box.south west) rectangle (current bounding box.north east);
\end{tikzpicture}
}\!\!,\hbox{\begin{tikzpicture} [scale=1.2,transform shape] %% DO NOT CHANGE

\def\deltax{0.3} %% CAN BE CHANGED
\def\deltay{0.5} %% DO NOT CHANGE

%\path[use as bounding box] (-\deltax,-\deltay) rectangle (\deltax,\deltay);

\node [dot, fill=\Zbwcolour] (mult) at (0,0) {};
%\node (mult_label_out) at (0,+\deltay) {};
%\draw[-] (mult) to (mult_label_out);

%\draw (current bounding box.south west) rectangle (current bounding box.north east);
\end{tikzpicture}
}\!\!)$ on $\starComplexs[\frac{1}{\omega_{uv}} \starIntegersModPow{2\omega+1}{3}]$. Time states $\ket{t}$ are chosen to have square norm $\omega_{uv}$, and energy states $\ket{E}$ are chosen to have square norm $\omega_{ir}$. Having said this, we define the following unitary module for $\hbox{\begin{tikzpicture} [scale=1.2,transform shape] %% DO NOT CHANGE

\def\deltax{0.3} %% CAN BE CHANGED
\def\deltay{0.5} %% DO NOT CHANGE

%\path[use as bounding box] (-\deltax,-\deltay) rectangle (\deltax,\deltay);

\node [dot, fill=\Dcolour] (mult) at (0,0) {};
%\node (mult_label_out) at (0,+\deltay) {};
%\draw[-] (mult) to (mult_label_out);

%\draw (current bounding box.south west) rectangle (current bounding box.north east);
\end{tikzpicture}
}\!\!$, a coherent version of the unitary representation from above:
\begin{equation}
	\begin{tikzpicture}
	\begin{pgfonlayer}{nodelayer}
		\node [style=box,draw=red,text=red,minimum height=10mm] (Ubox) at (0,0) {$U$};
		\node [style=none,text=red,inner sep = 2pt] (inr) at (-2.2,+0.5) {};
		\node [style=none] (box_inr) at (2,+0.5) {};
		\node [style=none,text=red,inner sep = 3pt] (inrr) at (-2.2,-0.5) {};
		\node [style=none] (box_inrr) at (0,-0.5) {};
	\end{pgfonlayer}
	\begin{pgfonlayer}{edgelayer}
		\draw[-,draw=red] (inr) to (box_inr);
		\draw[-,draw=red] (inrr) to (box_inrr);
	\end{pgfonlayer}
\end{tikzpicture}
	\hspace{2mm}
	:= 
	\hspace{2mm}
	\sum_{t} \frac{1}{\omega_{uv}} 
	\begin{tikzpicture}
	\begin{pgfonlayer}{nodelayer}
		\node [style=box,draw=red,text=red,minimum height=6mm] (Ubox) at (1,0.5) {$U_t$};
		\node [style=none,text=red,inner sep = 2pt] (inr) at (-2.2,+0.5) {};
		\node [style=none] (box_inr) at (3.5,+0.5) {};
		\node [style=none,text=red,inner sep = 3pt] (inrr) at (-2.2,-0.5) {};
		\node [style=effect,draw=red,text=red] (box_inrr) at (-0.5,-0.5) {$\delta_t$};
	\end{pgfonlayer}
	\begin{pgfonlayer}{edgelayer}
		\draw[-,draw=red] (inr) to (box_inr);
		\draw[-,draw=red] (inrr) to (box_inrr);
	\end{pgfonlayer}
\end{tikzpicture}
\end{equation}
The unitary module above is diagonalised by $\hbox{\begin{tikzpicture} [scale=1.2,transform shape] %% DO NOT CHANGE

\def\deltax{0.3} %% CAN BE CHANGED
\def\deltay{0.5} %% DO NOT CHANGE

%\path[use as bounding box] (-\deltax,-\deltay) rectangle (\deltax,\deltay);

\node [dot, fill=\Zbwcolour] (mult) at (0,0) {};
%\node (mult_label_out) at (0,+\deltay) {};
%\draw[-] (mult) to (mult_label_out);

%\draw (current bounding box.south west) rectangle (current bounding box.north east);
\end{tikzpicture}
}\!\!$, in the following sense:
\begin{equation}\label{timeTranslationMomentumRepInternalised}
	\begin{tikzpicture}
	\begin{pgfonlayer}{nodelayer}
		\node [style=none] (in) at (-2,+0.5) {};
		\node [style=box,draw=red,text=red,minimum height=10mm] (Ubox) at (0,0) {$U^\dagger$};
		\node [style=none] (out) at (2,+0.5) {};
		\node [style=none] (box_outr) at (0,-0.5) {};
		\node [style=none] (outr) at (2,-0.5) {};
	\end{pgfonlayer}
	\begin{pgfonlayer}{edgelayer}
		\draw[-,draw=red] (in) to (out);
		\draw[-,draw=red] (box_outr) to (outr);
	\end{pgfonlayer}
\end{tikzpicture}
	\hspace{2mm}
	=
	\hspace{2mm}
	\begin{tikzpicture}
	\begin{pgfonlayer}{nodelayer}
		\node [style=Zbwdot,draw=red] (dot) at (-2,0) {};
		\node [style=none] (dot_outl) at (-0.5,+0.5) {};
		\node [style=none] (out) at (2,+0.5){};
		\node [style=box,draw=red,text=red,minimum height=8mm] (Ubox) at (0,-0.5) {$U^\dagger$};
		\node [style=none] (in) at (-3.5,0) {};
		\node [style=none] (box_outl) at (0,0) {};
		\node [style=Zbwdot,draw=red] (outl) at (1.25,0) {};
		\node [style=none] (box_outr) at (0,-0.5) {};
		\node [style=none] (outr) at (2,-0.5) {};
	\end{pgfonlayer}
	\begin{pgfonlayer}{edgelayer}
		\draw[-,draw=red] (in) to (dot);
		\draw[-,draw=red,out=-45,in=180] (dot) to (Ubox);
		\draw[-,draw=red,out=+45,in=180] (dot) to (dot_outl.center);
		\draw[-,draw=red] (dot_outl.center) to (out);
		\draw[-,draw=red] (box_outl) to (outl);
		\draw[-,draw=red] (box_outr) to (outr);
	\end{pgfonlayer}
\end{tikzpicture}
\end{equation}

The reason why we can work on the controlling parameter space $\starComplexs[\frac{1}{\omega_{ir}} \starIntegersModPow{2\omega+1}{3}]$, rather than the full field space $\SpaceH^{\otimes\frac{1}{\omega_{ir}}\starIntegersModPow{2\omega+1}{3}}$, is that the ladder operators take the following form in the Heisenberg picture:
\begin{equation}
	e^{i\frac{2\pi}{h} H t}\,a^\dagger(\underline{p})\,e^{-i\frac{2\pi}{h} H t} 
	= 
	e^{+iE_{\underline{p}}t} a^\dagger(\underline{p})
	\hspace{2cm}
	e^{i\frac{2\pi}{h} H t}\,a(\underline{p})\,e^{-i\frac{2\pi}{h} H t} 
	= 
	e^{-iE_{\underline{p}}t} a(\underline{p})
\end{equation}
This means that the Heisenberg picture version of the coherently controlled creation operator can be written as follows:
\begin{equation}\label{gammaMapConstructorHeisenberg}
	\begin{tikzpicture}
	\begin{pgfonlayer}{nodelayer}
		\node [style=box,minimum height = 18mm, minimum width = 8mm] (box) at (0, 0) {$\bar{\gamma}^\dagger$};
		\node [style=none] (inl) at (-2.2,1) {};
		\node [style=none] (box_inl) at (0,1) {};
		\node [style=none,text=red,inner sep = 3pt] (inr) at (-2.2,-0.5) {$p$};
		\node [style=none] (box_inr) at (0,-0.5) {};
		\node [style=none,text=red,inner sep = 3pt] (inrr) at (-2.2,-1.5) {$t$};
		\node [style=none] (box_inrr) at (0,-1.5) {};
		\node [style=none] (out) at (2.2,0) {};
		\node [style=none] (box_out) at (0,0) {};
	\end{pgfonlayer}
	\begin{pgfonlayer}{edgelayer}
		\draw[-] (inl.center) to (box_inl); 
		\draw[-,draw=red] (inr) to (box_inr);
		\draw[-,draw=red] (inrr) to (box_inrr);
		\draw[-] (box_out) to (out.center);
	\end{pgfonlayer}
\end{tikzpicture}
	\hspace{3mm}
	=
	\hspace{3mm}
	\begin{tikzpicture}
	\begin{pgfonlayer}{nodelayer}
		\node [style=box,minimum height = 12mm, minimum width = 8mm] (box) at (2, 0.25) {$\gamma^\dagger$};
		\node [style=box,draw=red,text=red,minimum height=10mm] (Ubox) at (0,-1) {$U$};
		\node [style=none] (inl) at (-2.2,1) {};
		\node [style=none] (box_inl) at (2,1) {};
		\node [style=none,text=red,inner sep = 2pt] (inr) at (-2.2,-0.5) {$p$};
		\node [style=none] (box_inr) at (2,-0.5) {};
		\node [style=none,text=red,inner sep = 3pt] (inrr) at (-2.2,-1.5) {$t$};
		\node [style=none] (box_inrr) at (0,-1.5) {};
		\node [style=none] (out) at (4.2,0.25) {};
		\node [style=none] (box_out) at (2,0.25) {};
	\end{pgfonlayer}
	\begin{pgfonlayer}{edgelayer}
		\draw[-] (inl.center) to (box_inl); 
		\draw[-,draw=red] (inr) to (box_inr);
		\draw[-,draw=red] (inrr) to (box_inrr);
		\draw[-] (box_out) to (out.center);
	\end{pgfonlayer}
\end{tikzpicture}
\end{equation}
By applying this to position $\underline{x}$ and time $t$, we obtain the Lorentz invariant field operator $\phi(\underline{x},t)$ in the Heisenberg picture (writing $p \cdot x$ for $E_{\underline{p}}t-\underline{p}\cdot\underline{x}$):
\begin{align}
	\phi(\underline{x},t) 
	\hspace{2mm}
	&:= 
	\hspace{2mm}
	\phi^+(\underline{x},t) + \phi^-(\underline{x},t)
	\nonumber\\
	\phi^+(\underline{x},t) 
	\hspace{2mm}
	&:= 
	\hspace{2mm}
	\begin{tikzpicture}
	\begin{pgfonlayer}{nodelayer}
		\node [style=box,minimum height = 18mm, minimum width = 8mm] (box) at (0, 0) {$\bar{\gamma}$};
		\node [style=none] (in) at (-2.2,0) {};
		\node [style=none] (box_in) at (0,0) {};
		\node [style=none] (outl) at (2.2,1) {};
		\node [style=none] (box_outl) at (0,1) {};
		\node [style=none] (box_outr) at (0,-0.3) {};
		\node [style=none] (box_outrr) at (0,-1.5) {};
		\node [style=none] (outr) at (1.5,-0.3) {};
		\node [style=none] (outrr) at (1,-1.5) {};
		\node [style=Xbwdot,draw=red] (tdot) at (1.5,-2.25) {};
		\node [style=Xbwdot,draw=red] (xdot) at (2.5,-2.25) {};
		\node [style=none] (box_outr_r) at (1,-3) {};
		\node [style=none] (box_outrr_r) at (1.5,-4) {};
		\node [style=state,draw=red,text=red,minimum height=0mm] (inr) at (-2.2,-1.8) {$\delta_{\underline{x}}$};
		% \node [style=none] (box_inr) at (0,-0.3) {};
		\node [style=state,draw=red,text=red,minimum height=0mm] (inrr) at (-2.2,-3) {$\delta_t$};
		% \node [style=none] (out) at (2.2,0) {};
		% \node [style=none] (box_out) at (0,0) {};
	\end{pgfonlayer}
	\begin{pgfonlayer}{edgelayer}
		\draw[-] (in.center) to (box_in);
		\draw[-] (box_outl) to (outl.center);
		\draw[-,draw=red] (box_outr) to (outr.center);
		\draw[-,draw=red] (box_outrr) to (outrr.center);
		\draw[-,draw=red,out=0,in=90] (outr.center) to (xdot);
		\draw[-,draw=red,out=0,in=90] (outrr.center) to (tdot);
		\draw[-,draw=red,out=0,in=-90] (box_outr_r.center) to (tdot);
		\draw[-,draw=red,out=0,in=-90] (box_outrr_r.center) to (xdot);
		\draw[-,draw=red,out=0,in=180] (inr) to (box_outrr_r.center);
		\draw[-,draw=red,out=0,in=180] (inrr) to (box_outr_r.center);
	\end{pgfonlayer}
\end{tikzpicture}
	\hspace{2mm}
	=
	\hspace{2mm}
	\sum_{\underline{p}} \frac{1}{\omega_{ir}^3} \frac{1}{2E_{\underline{p}}} a({p})e^{-i2\pi p\cdot x}
	\nonumber \\
	\phi^-(\underline{x},t) 
	\hspace{2mm}
	&:= 
	\hspace{2mm}
	\begin{tikzpicture}
	\begin{pgfonlayer}{nodelayer}
		\node [style=box,minimum height = 18mm, minimum width = 8mm] (box) at (0, 0) {$\bar{\gamma}^\dagger$};
		\node [style=none] (inl) at (-2.2,1) {};
		\node [style=none] (box_inl) at (0,1) {};
		\node [style=state,draw=red,text=red,minimum height=0mm] (inr) at (-2.2,-0.3) {$\delta_{\underline{x}}$};
		\node [style=none] (box_inr) at (0,-0.3) {};
		\node [style=state,draw=red,text=red,minimum height=0mm] (inrr) at (-2.2,-1.5) {$\delta_t$};
		\node [style=none] (box_inrr) at (0,-1.5) {};
		\node [style=none] (out) at (2.2,0) {};
		\node [style=none] (box_out) at (0,0) {};
	\end{pgfonlayer}
	\begin{pgfonlayer}{edgelayer}
		\draw[-] (inl.center) to (box_inl); 
		\draw[-,draw=red] (inr) to (box_inr);
		\draw[-,draw=red] (inrr) to (box_inrr);
		\draw[-] (box_out) to (out.center);
	\end{pgfonlayer}
\end{tikzpicture}
	\hspace{2mm}
	=
	\hspace{2mm}
	\sum_{\underline{p}} \frac{1}{\omega_{ir}^3} \frac{1}{2E_{\underline{p}}} a^\dagger({p})e^{+i2\pi p\cdot x}
	\nonumber \\
%	\text{$\phi^+$, $\phi^-$ and then $\phi=\phi^+ + \phi^-$ on a third line.}
\end{align}
In the appendix, we provide a diagrammatic proof that the following commutation relation holds between the coherently controlled creation/destruction operators in the Heisenberg picture:
\begin{equation}
	\begin{tikzpicture}
	\begin{pgfonlayer}{nodelayer}
		\node [style=box,minimum height = 18mm, minimum width = 8mm] (box) at (0, 0) {$\bar{\gamma}^\dagger$};
		\node [style=none] (inl) at (-2.2,1) {};
		\node [style=none] (box_inl) at (0,1) {};
		\node [style=none,draw=red,text=red,minimum height=0mm] (inr) at (-2.2,-0.3) {};
		\node [style=none] (box_inr) at (0,-0.3) {};
		\node [style=none,draw=red,text=red,minimum height=0mm] (inrr) at (-2.2,-1.5) {};
		\node [style=none] (box_inrr) at (0,-1.5) {};
		\node [style=none] (out) at (2.2,0) {};
		\node [style=none] (box_out) at (0,0) {};
	\end{pgfonlayer}
	\begin{pgfonlayer}{edgelayer}
		\draw[-] (inl.center) to (box_inl); 
		\draw[-,draw=red] (inr) to (box_inr);
		\draw[-,draw=red] (inrr) to (box_inrr);
		\draw[-] (box_out) to (out.center);
	\end{pgfonlayer}

	\begin{pgfonlayer}{nodelayer}
		\node [style=box,minimum height = 18mm, minimum width = 8mm] (box) at (3, 0) {$\bar{\gamma}$};
		\node [style=none] (outl) at (5.4,1) {};
		\node [style=none] (box_outl) at (3,1) {};
		\node [style=none] (box_outr) at (3,-0.3) {};
		\node [style=none] (box_outrr) at (3,-1.5) {};
		\node [style=none] (outr) at (4.5,-0.3) {};
		\node [style=none] (outrr) at (4,-1.5) {};
		\node [style=Xbwdot,draw=red] (tdot) at (4.5,-2.25) {};
		\node [style=Xbwdot,draw=red] (xdot) at (5.5,-2.25) {};
		\node [style=none] (box_outr_r) at (4,-3) {};
		\node [style=none] (box_outrr_r) at (4.5,-4) {};
		\node [style=none,draw=red,text=red,minimum height=0mm] (inr) at (-2.2,-2.7) {};
		% \node [style=none] (box_inr) at (0,-0.3) {};
		\node [style=none,draw=red,text=red,minimum height=0mm] (inrr) at (-2.2,-3.9) {};
		% \node [style=none] (out) at (2.2,0) {};
		% \node [style=none] (box_out) at (0,0) {};
	\end{pgfonlayer}
	\begin{pgfonlayer}{edgelayer}
		\draw[-] (box_outl) to (outl.center);
		\draw[-,draw=red] (box_outr) to (outr.center);
		\draw[-,draw=red] (box_outrr) to (outrr.center);
		\draw[-,draw=red,out=0,in=90] (outr.center) to (xdot);
		\draw[-,draw=red,out=0,in=90] (outrr.center) to (tdot);
		\draw[-,draw=red,out=0,in=-90] (box_outr_r.center) to (tdot);
		\draw[-,draw=red,out=0,in=-90] (box_outrr_r.center) to (xdot);
		\draw[-,draw=red,out=0,in=180] (inr) to (box_outrr_r.center);
		\draw[-,draw=red,out=0,in=180] (inrr) to (box_outr_r.center);
	\end{pgfonlayer}
\end{tikzpicture}
	\hspace{2mm}
	-
	\hspace{2mm}
	\begin{tikzpicture}
	\begin{pgfonlayer}{nodelayer}
		\node [style=box,minimum height = 18mm, minimum width = 8mm] (box) at (6, 0) {$\bar{\gamma}^\dagger$};
		\node [style=none] (inl) at (2.2,1) {};
		\node [style=none] (box_inl) at (6,1) {};
		\node [style=none,draw=red,text=red,minimum height=0mm] (inr) at (-3,-1) {};
		\node [style=none] (box_inr) at (5.5,-0.3) {};
		\node [style=none,draw=red,text=red,minimum height=0mm] (inrr) at (-3,-2) {};
		\node [style=none] (box_inrr) at (5.5,-1.5) {};
		\node [style=none] (out) at (8.2,0) {};
		\node [style=none] (box_out) at (6,0) {};
		\node [style=none] (inr2) at (2.5,-3.5) {};
		\node [style=none] (inrr2) at (2.5,-4.5) {};
	\end{pgfonlayer}
	\begin{pgfonlayer}{edgelayer}
		\draw[-] (inl.center) to (box_inl); 
		\draw[-,out=0,in=180,draw=red] (inr) to (inr2.center);
		\draw[-,out=0,in=180,draw=red] (inrr) to (inrr2.center);
		\draw[-,out=0,in=180,draw=red] (inr2.center) to (box_inr);
		\draw[-,out=0,in=180,draw=red] (inrr2.center) to (box_inrr);
		\draw[-] (box_out) to (out.center);
	\end{pgfonlayer}
	\begin{pgfonlayer}{nodelayer}
		\node [style=box,minimum height = 18mm, minimum width = 8mm] (box) at (0, 0) {$\bar{\gamma}$};
		\node [style=none] (in) at (-3,0) {};
		\node [style=none] (box_in) at (0,0) {};
		\node [style=none] (outl) at (2.2,1) {};
		\node [style=none] (box_outl) at (0,1) {};
		\node [style=none] (box_outr) at (0,-0.3) {};
		\node [style=none] (box_outrr) at (0,-1.5) {};
		\node [style=none] (outr) at (1.5,-0.3) {};
		\node [style=none] (outrr) at (1,-1.5) {};
		\node [style=Xbwdot,draw=red] (tdot) at (1.5,-2.25) {};
		\node [style=Xbwdot,draw=red] (xdot) at (2.5,-2.25) {};
		\node [style=none] (box_outr_r) at (1,-3) {};
		\node [style=none] (box_outrr_r) at (1.5,-4) {};
		\node [style=none,draw=red,text=red,minimum height=0mm] (inr) at (-3,-3) {};
		% \node [style=none] (box_inr) at (0,-0.3) {};
		\node [style=none,draw=red,text=red,minimum height=0mm] (inrr) at (-3,-4) {};
		% \node [style=none] (out) at (2.2,0) {};
		% \node [style=none] (box_out) at (0,0) {};
	\end{pgfonlayer}
	\begin{pgfonlayer}{edgelayer}
		\draw[-] (in.center) to (box_in);
		\draw[-] (box_outl) to (outl.center);
		\draw[-,draw=red] (box_outr) to (outr.center);
		\draw[-,draw=red] (box_outrr) to (outrr.center);
		\draw[-,draw=red,out=0,in=90] (outr.center) to (xdot);
		\draw[-,draw=red,out=0,in=90] (outrr.center) to (tdot);
		\draw[-,draw=red,out=0,in=-90] (box_outr_r.center) to (tdot);
		\draw[-,draw=white,line width=3pt,out=0,in=-90] (box_outrr_r.center) to (xdot);
		\draw[-,draw=red,out=0,in=-90] (box_outrr_r.center) to (xdot);
		\draw[-,draw=white,line width=3pt,out=0,in=180] (inr) to (box_outrr_r.center);
		\draw[-,draw=white,line width=3pt,out=0,in=180] (inrr) to (box_outr_r.center);
		\draw[-,draw=red,out=0,in=180] (inr) to (box_outrr_r.center);
		\draw[-,draw=red,out=0,in=180] (inrr) to (box_outr_r.center);
	\end{pgfonlayer}
\end{tikzpicture}
	\hspace{3mm}
	=_\tau
	\hspace{3mm}
	\raisebox{-7mm}{\begin{tikzpicture}
	\begin{pgfonlayer}{nodelayer}
		\node[style=none] (in) at (0,+2.5) {};
		\node[style=none] (out) at (9,+2.5) {};
		\node[style=none,draw=red,text=red,minimum height=0mm] (iny) at (0,+1.25) {};
		\node[style=none,draw=red,text=red,minimum height=0mm] (iny0) at (0,0) {};
		\node[style=none,draw=red,text=red,minimum height=0mm] (inx) at (0,-1.25) {};
		\node[style=none,draw=red,text=red,minimum height=0mm] (inx0) at (0,-2.5) {};
		\node[style=none] (idy) at (2,+1.25) {};
		\node[style=none] (idy0) at (2,0) {};
		\node[style=boxsmall,draw=red] (antipodex) at (2,-1.25) {};
		\node[style=boxsmall,draw=red] (antipodex0) at (2,-2.5) {};
		\node[style=Zbwdot,draw=red] (posadd) at (4,0) {}; 
		\node[style=Zbwdot,draw=red] (timeadd) at (4,-1.25) {}; 
		\node[style=Zbwdot,draw=red] (unit) at (5,-0.625) {};
		\node[style=none] (Uinl) at (6.5,-0.625) {};
		\node[style=none] (Uinr) at (6.5,-1.25) {};
		\node[style=box,draw=red,text=red,minimum height=6mm,minimum width=6mm] (U) at (6.5,-0.9375) {$U$}; 
		\node[style=none] (Uout) at (6.5,-1.25) {};
		\node[style=none] (dotinl) at (6.5,0) {};
		\node[style=Zbwdot,draw=red] (dot) at (7.6,-0.625) {};
	\end{pgfonlayer}
	\begin{pgfonlayer}{edgelayer}
		\draw[-] (in) to (out);
		\draw[-,draw=red] (iny) to (idy.center);
		\draw[-,draw=red] (iny0) to (idy0.center);
		\draw[-,draw=red] (inx) to (antipodex);
		\draw[-,draw=red] (inx0) to (antipodex0);
		\draw[-,draw=red,out=0,in=+135] (idy.center) to (posadd);
		\draw[-,draw=red,out=0,in=+135] (idy0.center) to (timeadd);
		\draw[-,draw=red,out=0,in=-135] (antipodex) to (posadd);
		\draw[-,draw=red,out=0,in=-135] (antipodex0) to (timeadd);
		\draw[-,draw=red] (posadd) to (dotinl.center);
		\draw[-,draw=red] (unit) to (Uinl);
		\draw[-,draw=red] (timeadd) to (Uinr.center);
		\draw[-,draw=red,out=0,in=+90] (dotinl.center) to (dot);
		\draw[-,draw=red,out=0,in=-90] (Uinr.center) to (dot);

	\end{pgfonlayer}
\end{tikzpicture}}
\end{equation}
By evaluating the commutation relation on two 4-position states, we finally see that the propagator $D(x-y) = [\phi^+(x),\phi^-(y)]$ takes the following form:
\begin{equation}
	D(x-y)
	\hspace{2mm}
	=_\tau
	\hspace{2mm}
	\begin{tikzpicture}
	\begin{pgfonlayer}{nodelayer}
		\node[style=none] (in) at (-2,+2.5) {};
		\node[style=none] (out) at (9,+2.5) {};
		\node[style=state,draw=red,text=red,minimum height=0mm] (iny) at (0,+1.25) {$\delta_{\underline{y}}$};
		\node[style=state,draw=red,text=red,minimum height=0mm] (iny0) at (0,0) {$\delta_{y_0}$};
		\node[style=state,draw=red,text=red,minimum height=0mm] (inx) at (0,-1.25) {$\delta_{\underline{x}}$};
		\node[style=state,draw=red,text=red,minimum height=0mm] (inx0) at (0,-2.5) {$\delta_{x_0}$};
		\node[style=none] (idy) at (2,+1.25) {};
		\node[style=none] (idy0) at (2,0) {};
		\node[style=boxsmall,draw=red] (antipodex) at (2,-1.25) {};
		\node[style=boxsmall,draw=red] (antipodex0) at (2,-2.5) {};
		\node[style=Zbwdot,draw=red] (posadd) at (4,0) {}; 
		\node[style=Zbwdot,draw=red] (timeadd) at (4,-1.25) {}; 
		\node[style=Zbwdot,draw=red] (unit) at (5,-0.625) {};
		\node[style=none] (Uinl) at (6.5,-0.625) {};
		\node[style=none] (Uinr) at (6.5,-1.25) {};
		\node[style=box,draw=red,text=red,minimum height=6mm,minimum width=6mm] (U) at (6.5,-0.9375) {$U$}; 
		\node[style=none] (Uout) at (6.5,-1.25) {};
		\node[style=none] (dotinl) at (6.5,0) {};
		\node[style=Zbwdot,draw=red] (dot) at (7.6,-0.625) {};
	\end{pgfonlayer}
	\begin{pgfonlayer}{edgelayer}
		\draw[-] (in) to (out);
		\draw[-,draw=red] (iny) to (idy.center);
		\draw[-,draw=red] (iny0) to (idy0.center);
		\draw[-,draw=red] (inx) to (antipodex);
		\draw[-,draw=red] (inx0) to (antipodex0);
		\draw[-,draw=red,out=0,in=+135] (idy.center) to (posadd);
		\draw[-,draw=red,out=0,in=+135] (idy0.center) to (timeadd);
		\draw[-,draw=red,out=0,in=-135] (antipodex) to (posadd);
		\draw[-,draw=red,out=0,in=-135] (antipodex0) to (timeadd);
		\draw[-,draw=red] (posadd) to (dotinl.center);
		\draw[-,draw=red] (unit) to (Uinl);
		\draw[-,draw=red] (timeadd) to (Uinr.center);
		\draw[-,draw=red,out=0,in=+90] (dotinl.center) to (dot);
		\draw[-,draw=red,out=0,in=-90] (Uinr.center) to (dot);

	\end{pgfonlayer}
\end{tikzpicture}
	\hspace{2mm}
	=
	\hspace{2mm}
	\sum_{\underline{p}} \frac{1}{\omega_{ir}^3} \frac{1}{2E_{\underline{p}}} e^{-i2\pi p\cdot(x-y)} \id{}
\end{equation}

\section{Conclusions and Future Work}
\label{section_conclusions}

We have shown how the fundamental building blocks of quantum field theory can be formulated within the framework of categorical quantum mechanics, by using well understood structures such as Frobenius algebras and modules. Unfortunately, the space available here is only enough to barely scratch the surface, and a number of interesting applications must be deferred to future work: these include the formulation of the interaction picture, Feynman diagrams and renormalisation, as well as the working out of interesting concrete examples such as $\phi^4$ theory and Yukawa theory. Somewhat further down the line, we foresee applications of the framework introduced here to a categorical formulation of spinors and Gauge boson, and subsequent tackling of interesting real-world theories such as QED. 

\newpage
\bibliographystyle{eptcs}
\bibliography{biblio}
%\nocite{*}
%\input{biblio.bbl}

\appendix

\newpage
\section{Graphical derivation of the propagator}

Here is a fully graphical derivation of the commutation relation between the coherently controlled constructor/destructor operators in the Heisenberg picture:
\begin{align}
	&\begin{tikzpicture}
	\begin{pgfonlayer}{nodelayer}
		\node [style=box,minimum height = 18mm, minimum width = 8mm] (box) at (0, 0) {$\bar{\gamma}^\dagger$};
		\node [style=none] (inl) at (-2.2,1) {};
		\node [style=none] (box_inl) at (0,1) {};
		\node [style=none,draw=red,text=red,minimum height=0mm] (inr) at (-2.2,-0.3) {};
		\node [style=none] (box_inr) at (0,-0.3) {};
		\node [style=none,draw=red,text=red,minimum height=0mm] (inrr) at (-2.2,-1.5) {};
		\node [style=none] (box_inrr) at (0,-1.5) {};
		\node [style=none] (out) at (2.2,0) {};
		\node [style=none] (box_out) at (0,0) {};
	\end{pgfonlayer}
	\begin{pgfonlayer}{edgelayer}
		\draw[-] (inl.center) to (box_inl); 
		\draw[-,draw=red] (inr) to (box_inr);
		\draw[-,draw=red] (inrr) to (box_inrr);
		\draw[-] (box_out) to (out.center);
	\end{pgfonlayer}

	\begin{pgfonlayer}{nodelayer}
		\node [style=box,minimum height = 18mm, minimum width = 8mm] (box) at (3, 0) {$\bar{\gamma}$};
		\node [style=none] (outl) at (5.4,1) {};
		\node [style=none] (box_outl) at (3,1) {};
		\node [style=none] (box_outr) at (3,-0.3) {};
		\node [style=none] (box_outrr) at (3,-1.5) {};
		\node [style=none] (outr) at (4.5,-0.3) {};
		\node [style=none] (outrr) at (4,-1.5) {};
		\node [style=Xbwdot,draw=red] (tdot) at (4.5,-2.25) {};
		\node [style=Xbwdot,draw=red] (xdot) at (5.5,-2.25) {};
		\node [style=none] (box_outr_r) at (4,-3) {};
		\node [style=none] (box_outrr_r) at (4.5,-4) {};
		\node [style=none,draw=red,text=red,minimum height=0mm] (inr) at (-2.2,-2.7) {};
		% \node [style=none] (box_inr) at (0,-0.3) {};
		\node [style=none,draw=red,text=red,minimum height=0mm] (inrr) at (-2.2,-3.9) {};
		% \node [style=none] (out) at (2.2,0) {};
		% \node [style=none] (box_out) at (0,0) {};
	\end{pgfonlayer}
	\begin{pgfonlayer}{edgelayer}
		\draw[-] (box_outl) to (outl.center);
		\draw[-,draw=red] (box_outr) to (outr.center);
		\draw[-,draw=red] (box_outrr) to (outrr.center);
		\draw[-,draw=red,out=0,in=90] (outr.center) to (xdot);
		\draw[-,draw=red,out=0,in=90] (outrr.center) to (tdot);
		\draw[-,draw=red,out=0,in=-90] (box_outr_r.center) to (tdot);
		\draw[-,draw=red,out=0,in=-90] (box_outrr_r.center) to (xdot);
		\draw[-,draw=red,out=0,in=180] (inr) to (box_outrr_r.center);
		\draw[-,draw=red,out=0,in=180] (inrr) to (box_outr_r.center);
	\end{pgfonlayer}
\end{tikzpicture}
	\hspace{2mm}
	-
	\hspace{2mm}
	\begin{tikzpicture}
	\begin{pgfonlayer}{nodelayer}
		\node [style=box,minimum height = 18mm, minimum width = 8mm] (box) at (6, 0) {$\bar{\gamma}^\dagger$};
		\node [style=none] (inl) at (2.2,1) {};
		\node [style=none] (box_inl) at (6,1) {};
		\node [style=none,draw=red,text=red,minimum height=0mm] (inr) at (-3,-1) {};
		\node [style=none] (box_inr) at (5.5,-0.3) {};
		\node [style=none,draw=red,text=red,minimum height=0mm] (inrr) at (-3,-2) {};
		\node [style=none] (box_inrr) at (5.5,-1.5) {};
		\node [style=none] (out) at (8.2,0) {};
		\node [style=none] (box_out) at (6,0) {};
		\node [style=none] (inr2) at (2.5,-3.5) {};
		\node [style=none] (inrr2) at (2.5,-4.5) {};
	\end{pgfonlayer}
	\begin{pgfonlayer}{edgelayer}
		\draw[-] (inl.center) to (box_inl); 
		\draw[-,out=0,in=180,draw=red] (inr) to (inr2.center);
		\draw[-,out=0,in=180,draw=red] (inrr) to (inrr2.center);
		\draw[-,out=0,in=180,draw=red] (inr2.center) to (box_inr);
		\draw[-,out=0,in=180,draw=red] (inrr2.center) to (box_inrr);
		\draw[-] (box_out) to (out.center);
	\end{pgfonlayer}
	\begin{pgfonlayer}{nodelayer}
		\node [style=box,minimum height = 18mm, minimum width = 8mm] (box) at (0, 0) {$\bar{\gamma}$};
		\node [style=none] (in) at (-3,0) {};
		\node [style=none] (box_in) at (0,0) {};
		\node [style=none] (outl) at (2.2,1) {};
		\node [style=none] (box_outl) at (0,1) {};
		\node [style=none] (box_outr) at (0,-0.3) {};
		\node [style=none] (box_outrr) at (0,-1.5) {};
		\node [style=none] (outr) at (1.5,-0.3) {};
		\node [style=none] (outrr) at (1,-1.5) {};
		\node [style=Xbwdot,draw=red] (tdot) at (1.5,-2.25) {};
		\node [style=Xbwdot,draw=red] (xdot) at (2.5,-2.25) {};
		\node [style=none] (box_outr_r) at (1,-3) {};
		\node [style=none] (box_outrr_r) at (1.5,-4) {};
		\node [style=none,draw=red,text=red,minimum height=0mm] (inr) at (-3,-3) {};
		% \node [style=none] (box_inr) at (0,-0.3) {};
		\node [style=none,draw=red,text=red,minimum height=0mm] (inrr) at (-3,-4) {};
		% \node [style=none] (out) at (2.2,0) {};
		% \node [style=none] (box_out) at (0,0) {};
	\end{pgfonlayer}
	\begin{pgfonlayer}{edgelayer}
		\draw[-] (in.center) to (box_in);
		\draw[-] (box_outl) to (outl.center);
		\draw[-,draw=red] (box_outr) to (outr.center);
		\draw[-,draw=red] (box_outrr) to (outrr.center);
		\draw[-,draw=red,out=0,in=90] (outr.center) to (xdot);
		\draw[-,draw=red,out=0,in=90] (outrr.center) to (tdot);
		\draw[-,draw=red,out=0,in=-90] (box_outr_r.center) to (tdot);
		\draw[-,draw=white,line width=3pt,out=0,in=-90] (box_outrr_r.center) to (xdot);
		\draw[-,draw=red,out=0,in=-90] (box_outrr_r.center) to (xdot);
		\draw[-,draw=white,line width=3pt,out=0,in=180] (inr) to (box_outrr_r.center);
		\draw[-,draw=white,line width=3pt,out=0,in=180] (inrr) to (box_outr_r.center);
		\draw[-,draw=red,out=0,in=180] (inr) to (box_outrr_r.center);
		\draw[-,draw=red,out=0,in=180] (inrr) to (box_outr_r.center);
	\end{pgfonlayer}
\end{tikzpicture}
	\hspace{3mm} =
	\nonumber \\
	= \hspace{3mm}&
	\scalebox{0.95}{$\begin{tikzpicture}

	\begin{pgfonlayer}{nodelayer}
		\node[style=none] (inl) at (-9,0.6) {};
		\node[style=none] (inr) at (-9,-0.6) {};
		\node[style=none] (inrr) at (-9,-1.1) {};
		\node[style=none] (outl) at (-5.5,0.6) {};
		\node[style=none] (outr) at (-5.5,-0.6) {};
		\node[style=box,draw=red,text=red] (U) at (-7,-0.9) {$U$};
		\node[style=none] (outrr) at (-7,-1.1) {};
		\node[style=Zbwdot,draw=red] (dot) at (-6,-2.1) {};
		\node (dotl) at (-5.5,-1.6) {};
		\node (dotr) at (-5.5,-2.6) {};
		\node (rightmost_dotr) at (16.5,-2.6) {};
	\end{pgfonlayer}
	\begin{pgfonlayer}{edgelayer}
		\draw[-] (inl) to (outl); 
		\draw[-,draw=red] (inr) to (outr); 
		\draw[-,draw=red] (inrr) to (outrr); 
		\draw[-,out=90,in=180,draw=red] (dot) to (dotl.center);
		\draw[-,out=-90,in=180,draw=red] (dot) to (dotr.center);
		\draw[-,draw=red] (dotr.center) to (rightmost_dotr.center);
	\end{pgfonlayer}
	
	\begin{pgfonlayer}{nodelayer}
		\node (lb) at (-4.5,-2) {};
		\node (lt) at (-4.5,2) {};
	\end{pgfonlayer}
	\begin{pgfonlayer}{edgelayer}
		\draw[-,out=135,in=-135] (lb) to (lt); 
	\end{pgfonlayer}

	\begin{pgfonlayer}{nodelayer}
		\node [style=box,minimum height = 12mm, minimum width = 8mm] (gamma) at (-2, 0) {$\gamma^\dagger$};
		\node [style=box,minimum height = 12mm, minimum width = 8mm] (gammad) at (+2,0) {$\gamma$};

		\node [style=none] (inl) at (-4.5,0.6) {};
		\node [style=none] (inr) at (-4.5,-0.6) {};		
		\node [style=none] (inrr) at (-4.5,-1.6) {};
		\node [style=none] (gammainl) at (-2,0.6) {};
		\node [style=none] (gammainr) at (-2,-0.6) {};

		\node [style=none] (outl) at (4.5,0.6) {};
		\node [style=none] (dotinl) at (3,-0.6) {};
		\node [style=none] (gammadoutl) at (+2,0.6) {};
		\node [style=none] (gammadoutr) at (+2,-0.6) {};
		\node [style=none] (dotinr) at (3,-1.6) {};
		\node [style=dot,draw=red] (dot) at (3.5,-1.1){};
	\end{pgfonlayer}
	\begin{pgfonlayer}{edgelayer}
		\draw[-] (gamma) to (gammad); 
		\draw[-] (inl) to (gammainl); 
		\draw[-,draw=red] (inr) to (gammainr); 
		\draw[-] (gammadoutl) to (outl); 
		\draw[-,draw=red] (gammadoutr) to (dotinl.center); 
		\draw[-,out=0,in=90,draw=red] (dotinl.center) to (dot);
		\draw[-,out=0,in=-90,draw=red] (dotinr.center) to (dot);
		\draw[-,draw=red] (inrr) to (dotinr.center);
	\end{pgfonlayer}

	\node [style=none] (minus) at (6,0) {$-$};

	\begin{pgfonlayer}{nodelayer}
		\node [style=box,minimum height = 12mm, minimum width = 8mm] (gamma) at (9, 0) {$\gamma$};
		\node [style=box,minimum height = 12mm, minimum width = 8mm] (gammad) at (13,0) {$\gamma^\dagger$};
		\node [style=none] (gammal) at (9,0.6) {};
		\node [style=none] (gammadl) at (13,0.6) {};

		\node [style=none] (inl) at (7.5,0) {};
		\node [style=none] (inr) at (7.5,-0.6) {};	
		\node [style=none] (inr2) at (7.5,-0.6) {};	
		\node [style=none] (inr3) at (9,-2.1) {};	
		\node [style=none] (inr4) at (11,-2.1) {};		
		\node [style=none] (inr5) at (12,-0.6) {};	
		\node [style=none] (inrr) at (7.5,-1.6) {};
		\node [style=none] (gammainl) at (9,0) {};
		\node [style=none] (gammainr) at (13,-0.6) {};

		\node [style=none] (outl) at (15.5,0) {};
		\node [style=none] (dotinl) at (10,-0.6) {};
		\node [style=none] (gammadoutl) at (13,0) {};
		\node [style=none] (gammadoutr) at (9,-0.6) {};
		\node [style=none] (dotinr) at (10,-1.6) {};
		\node [style=dot,draw=red] (dot) at (10.5,-1.1){};
	\end{pgfonlayer}
	\begin{pgfonlayer}{edgelayer}
		\draw[-] (gammal) to (gammadl); 
		\draw[-] (inl) to (gammainl); 
		\draw[-,draw=red] (inr) to (inr2.center); 
		\draw[-,draw=red,out=0,in=180] (inr2.center) to (inr3.center); 
		\draw[-,draw=red,out=0,in=180] (inr3.center) to (inr4.center); 
		\draw[-,draw=red,out=0,in=180] (inr4.center) to (inr5.center); 
		\draw[-,draw=red,out=0,in=180] (inr5.center) to (gammainr); 
		% \draw[-,draw=red] (inr) to (gammainr); 
		\draw[-] (gammadoutl) to (outl); 
		\draw[-,draw=red] (gammadoutr) to (dotinl.center); 
		\draw[-,out=0,in=90,draw=red] (dotinl.center) to (dot);
		\draw[-,out=0,in=-90,draw=red] (dotinr.center) to (dot);
		\draw[-,draw=red] (inrr) to (dotinr.center);
	\end{pgfonlayer}

	\begin{pgfonlayer}{nodelayer}
		\node (lb) at (15.5,-2) {};
		\node (lt) at (15.5,2) {};
		\node (in) at (16.5,0) {};
		\node (out) at (20,0) {};
	\end{pgfonlayer}
	\begin{pgfonlayer}{edgelayer}
		\draw[-,out=45,in=-45] (lb) to (lt); 
		\draw[-] (in) to (out);
	\end{pgfonlayer}

	\begin{pgfonlayer}{nodelayer}
		\node[style=box,draw=red,text=red] (U) at (16.5,-2.6) {$U^\dagger$};
		\node[style=Xbwdot,draw=red] (dot1) at (17.5,-3.3) {};
		\node (dot1l) at (16.5,-2.9) {};
		\node (dot1r) at (16.5,-3.9) {};
		\node[style=Xbwdot,draw=red] (dot2) at (18.5,-3.4) {};
		\node (dot2l) at (16.5,-2.3) {};
		\node (dot2r) at (16.5,-4.5) {};
		\node (leftmost_dot1r) at (-7,-3.9) {};
		\node (leftmost_dot2r) at (-7,-4.5) {};
		\node (more_leftmost_dot1r) at (-9,-3.9) {};
		\node (more_leftmost_dot2r) at (-9,-4.5) {};
	\end{pgfonlayer}
	\begin{pgfonlayer}{edgelayer}
		\draw[-,out=90,in=0,draw=red] (dot1) to (dot1l.center);
		\draw[-,out=-90,in=0,draw=red] (dot1) to (dot1r.center);
		\draw[-,out=90,in=0,draw=red] (dot2) to (dot2l.center);
		\draw[-,out=-90,in=0,draw=red] (dot2) to (dot2r.center);
		\draw[-,draw=red] (dot1r.center) to (leftmost_dot1r.center);
		\draw[-,draw=red] (dot2r.center) to (leftmost_dot2r.center);
		\draw[-,out=0,in=180,draw=red] (more_leftmost_dot2r.center) to (leftmost_dot1r.center);
		\draw[-,out=0,in=180,draw=red] (more_leftmost_dot1r.center) to (leftmost_dot2r.center);
	\end{pgfonlayer}
	
\end{tikzpicture}$}
	\hspace{1mm}=_\tau
	\nonumber \\
	=_\tau \hspace{3mm}&
	\begin{tikzpicture}
	\node (empty) at (0,3) {};
	\node (empty) at (0,-3) {};

	\begin{pgfonlayer}{nodelayer}
		\node[style=none] (idin) at (-5, 1.5) {};
		\node[style=none] (idout) at (+5, 1.5) {};
		\node[style=box,draw=red,text=red] (U) at (-3.5,0) {$U$};
		\node[style=none] (Uinl1) at (-5,0.3) {};
		\node[style=none] (Uinl2) at (-3.5,0.3) {};
		\node[style=none] (Uinr1) at (-5,-0.3) {};
		\node[style=none] (Uinr2) at (-3.5,-0.3) {};
		\node[style=box,draw=red,text=red] (Udag) at (+2,0) {$U^\dagger$};
		\node[style=none] (Udagoutl2) at (3.5,0.3) {};
		\node[style=none] (Udagoutl1) at (2,0.3) {};
		\node[style=none] (Udagoutr2) at (2.5,-0.3) {};
		\node[style=none] (Udagoutr1) at (2,-0.3) {};
		\node[style=Zbwdot,draw=red] (zdot1) at (0,0.5) {};
		\node[style=Zbwdot,draw=red] (zdot2) at (-2,-0.5) {};
		\node[style=Xbwdot,draw=red] (xdot1) at (3.5,-1) {};
		\node[style=Xbwdot,draw=red] (xdot2) at (4.5,-1) {};
		\node[style=none] (xdot2l) at (3.5,-2.6) {};
		\node[style=none] (xdot1l) at (2.5,-1.7) {};
		\node[style=none] (xdot2ins) at (-3.5,-2.6) {};
		\node[style=none] (xdot1ins) at (-3.5,-1.7) {};
		\node[style=none] (xdot1in) at (-5,-2.6) {};
		\node[style=none] (xdot2in) at (-5,-1.7) {};
	\end{pgfonlayer}
	\begin{pgfonlayer}{edgelayer}
		\draw[-] (idin) to (idout); 
		\draw[-,draw=red] (Uinl1) to (Uinl2); 
		\draw[-,draw=red] (Uinr1) to (Uinr2); 
		\draw[-,draw=red] (Udagoutl1) to (Udagoutl2.center); 
		\draw[-,draw=red] (Udagoutr1) to (Udagoutr2.center); 
		\draw[-,draw=red,out=0,in=135] (U) to (zdot1); 
		\draw[-,draw=red,out=-135,in=45] (zdot1) to (zdot2); 
		\draw[-,draw=red,out=-45,in=180] (zdot2) to (Udag); 
		\draw[-,draw=red,out=0,in=90] (Udagoutl2.center) to (xdot2); 
		\draw[-,draw=red,out=0,in=90] (Udagoutr2.center) to (xdot1); 
		\draw[-,draw=red,out=0,in=-90] (xdot2l.center) to (xdot2); 
		\draw[-,draw=red,out=0,in=-90] (xdot1l.center) to (xdot1); 
		\draw[-,draw=red] (xdot2ins.center) to (xdot2l.center); 
		\draw[-,draw=red] (xdot1ins.center) to (xdot1l.center); 
		\draw[-,draw=red,out=0,in=180] (xdot2in.center) to (xdot2ins.center); 
		\draw[-,draw=red,out=0,in=180] (xdot1in.center) to (xdot1ins.center); 
	\end{pgfonlayer}

\end{tikzpicture}
	\hspace{3mm}= \hspace{3mm}
	\begin{tikzpicture}
	\node (empty) at (0,3) {};
	\node (empty) at (0,-3) {};

	\begin{pgfonlayer}{nodelayer}
		\node[style=none] (idin) at (-5, 1.5) {};
		\node[style=none] (idout) at (+5, 1.5) {};
		\node[style=box,draw=red,text=red] (U) at (-3.5,0) {$U$};
		\node[style=none] (Uinl1) at (-5,0.3) {};
		\node[style=none] (Uinl2) at (-3.5,0.3) {};
		\node[style=none] (Uinr1) at (-5,-0.3) {};
		\node[style=none] (Uinr2) at (-3.5,-0.3) {};
		\node[style=box,draw=red,text=red] (Udag) at (+2,0) {$U^\dagger$};
		\node[style=none] (Udagoutl2) at (3.5,0.3) {};
		\node[style=none] (Udagoutl1) at (2,0.3) {};
		\node[style=none] (Udagoutr2) at (2.5,-0.3) {};
		\node[style=none] (Udagoutr1) at (2,-0.3) {};
		\node[style=Xbwdot,draw=red] (xdot1) at (3.5,-1) {};
		\node[style=Xbwdot,draw=red] (xdot2) at (4.5,-1) {};
		\node[style=none] (xdot2l) at (3.5,-2.6) {};
		\node[style=none] (xdot1l) at (2.5,-1.7) {};
		\node[style=none] (xdot2ins) at (-3.5,-2.6) {};
		\node[style=none] (xdot1ins) at (-3.5,-1.7) {};
		\node[style=none] (xdot1in) at (-5,-2.6) {};
		\node[style=none] (xdot2in) at (-5,-1.7) {};
	\end{pgfonlayer}
	\begin{pgfonlayer}{edgelayer}
		\draw[-] (idin) to (idout); 
		\draw[-,draw=red] (Uinl1) to (Uinl2); 
		\draw[-,draw=red] (Uinr1) to (Uinr2); 
		\draw[-,draw=red] (Udagoutl1) to (Udagoutl2.center); 
		\draw[-,draw=red] (Udagoutr1) to (Udagoutr2.center); 
		\draw[-,draw=red,out=0,in=180] (U) to (Udag); 
		\draw[-,draw=red,out=0,in=90] (Udagoutl2.center) to (xdot2); 
		\draw[-,draw=red,out=0,in=90] (Udagoutr2.center) to (xdot1); 
		\draw[-,draw=red,out=0,in=-90] (xdot2l.center) to (xdot2); 
		\draw[-,draw=red,out=0,in=-90] (xdot1l.center) to (xdot1); 
		\draw[-,draw=red] (xdot2ins.center) to (xdot2l.center); 
		\draw[-,draw=red] (xdot1ins.center) to (xdot1l.center); 
		\draw[-,draw=red,out=0,in=180] (xdot2in.center) to (xdot2ins.center); 
		\draw[-,draw=red,out=0,in=180] (xdot1in.center) to (xdot1ins.center); 
	\end{pgfonlayer}

\end{tikzpicture}
	\hspace{3mm}=
	\nonumber \\
	= \hspace{3mm}&
	\begin{tikzpicture}
	\node (empty) at (0,3) {};
	\node (empty) at (0,-3) {};

	\begin{pgfonlayer}{nodelayer}
		\node[style=none] (idin) at (-5, 1.5) {};
		\node[style=none] (idout) at (+5, 1.5) {};
		\node[style=box,draw=red,text=red] (U) at (-3.5,0) {$U$};
		\node[style=none] (Uinl1) at (-5,0.3) {};
		\node[style=none] (Uinl2) at (-3.5,0.3) {};
		\node[style=none] (Uinr1) at (-5,-0.3) {};
		\node[style=none] (Uinr2) at (-3.5,-0.3) {};
		\node[style=box,draw=red,text=red] (Udag) at (+2,0) {$U$};
		\node[style=none] (Udagoutl2) at (3.5,0.3) {};
		\node[style=none] (Udagoutl1) at (2,0.3) {};
		\node[style=Xbwdot,draw=red] (xdot2) at (4.5,-0.7) {};
		\node[style=none] (xdot2l) at (3.5,-1.9) {};
		\node[style=none] (xdot1l) at (2,-0.3) {};
		\node[style=none] (xdot2ins) at (-3.5,-1.9) {};
		\node[style=boxsmall,draw=red] (xdot1ins) at (-3.5,-1.2) {};
		\node[style=none] (xdot1in) at (-5,-1.9) {};
		\node[style=none] (xdot2in) at (-5,-1.2) {};
	\end{pgfonlayer}
	\begin{pgfonlayer}{edgelayer}
		\draw[-] (idin) to (idout); 
		\draw[-,draw=red] (Uinl1) to (Uinl2); 
		\draw[-,draw=red] (Uinr1) to (Uinr2); 
		\draw[-,draw=red] (Udagoutl1) to (Udagoutl2.center); 
		\draw[-,draw=red] (Udagoutr1) to (Udagoutr2.center); 
		\draw[-,draw=red,out=0,in=180] (U) to (Udag); 
		\draw[-,draw=red,out=0,in=90] (Udagoutl2.center) to (xdot2); 
		\draw[-,draw=red,out=0,in=-90] (xdot2l.center) to (xdot2); 
		\draw[-,draw=red] (xdot2ins.center) to (xdot2l.center); 
		\draw[-,draw=red,out=0,in=180] (xdot1ins.center) to (xdot1l.center); 
		\draw[-,draw=red,out=0,in=180] (xdot2in.center) to (xdot2ins.center); 
		\draw[-,draw=red,out=0,in=180] (xdot1in.center) to (xdot1ins.center); 
	\end{pgfonlayer}

\end{tikzpicture}
	\hspace{3mm}= \hspace{3mm}
	\begin{tikzpicture}
	\node (empty) at (0,3) {};
	\node (empty) at (0,-3) {};

	\begin{pgfonlayer}{nodelayer}
		\node[style=none] (idin) at (-5, 1.5) {};
		\node[style=none] (idout) at (+5, 1.5) {};
		\node[style=none] (Uinl1) at (-5,0.3) {};
		\node[style=none] (Uinl2) at (2,0.3) {};
		\node[style=none] (Uinr1) at (-5,-0.3) {};
		\node[style=Zbwdot,draw=red] (mult) at (-1.5,-0.6) {};
		\node[style=box,draw=red,text=red] (Udag) at (+2,0) {$U$};
		\node[style=none] (Udagoutl2) at (3.5,0.3) {};
		\node[style=none] (Udagoutl1) at (2,0.3) {};
		\node[style=Xbwdot,draw=red] (xdot2) at (4.5,-0.7) {};
		\node[style=none] (xdot2l) at (3.5,-1.9) {};
		\node[style=none] (xdot1l) at (2,-0.3) {};
		\node[style=none] (xdot2ins) at (-3.5,-1.9) {};
		\node[style=boxsmall,draw=red] (xdot1ins) at (-3.5,-1.2) {};
		\node[style=none] (xdot1in) at (-5,-1.9) {};
		\node[style=none] (xdot2in) at (-5,-1.2) {};
	\end{pgfonlayer}
	\begin{pgfonlayer}{edgelayer}
		\draw[-] (idin) to (idout); 
		\draw[-,draw=red] (Uinl1) to (Uinl2); 
		\draw[-,draw=red,out=0,in=135] (Uinr1) to (mult); 
		\draw[-,draw=red] (Udagoutl1) to (Udagoutl2.center); 
		\draw[-,draw=red] (Udagoutr1) to (Udagoutr2.center); 
		\draw[-,draw=red,out=0,in=180] (mult) to (xdot1l); 
		\draw[-,draw=red,out=0,in=90] (Udagoutl2.center) to (xdot2); 
		\draw[-,draw=red,out=0,in=-90] (xdot2l.center) to (xdot2); 
		\draw[-,draw=red] (xdot2ins.center) to (xdot2l.center); 
		\draw[-,draw=red,out=0,in=-135] (xdot1ins.center) to (mult); 
		\draw[-,draw=red,out=0,in=180] (xdot2in.center) to (xdot2ins.center); 
		\draw[-,draw=red,out=0,in=180] (xdot1in.center) to (xdot1ins.center); 
	\end{pgfonlayer}

\end{tikzpicture}
	\hspace{3mm}=
	\nonumber \\
	= \hspace{3mm}&
	\begin{tikzpicture}
	\node (empty) at (0,3) {};
	\node (empty) at (0,-3) {};

	\begin{pgfonlayer}{nodelayer}
		\node[style=none] (idin) at (-5, 1.5) {};
		\node[style=none] (idout) at (+5, 1.5) {};
		\node[style=none] (Uinl1) at (-5,0.3) {};
		\node[style=none] (Uinl2) at (2,0.3) {};
		\node[style=none] (Uinr1) at (-5,-0.3) {};
		\node[style=Zbwdot,draw=red] (mult) at (-1.5,-0.6) {};
		\node[style=box,draw=red,text=red] (Udag) at (+2,0) {$U$};
		\node[style=none] (Udagoutl2) at (3.5,0.3) {};
		\node[style=none] (Udagoutl1) at (2,0.3) {};
		\node[style=Zbwdot,draw=red] (xdot2) at (4.5,-0.7) {};
		\node[style=none] (xdot2l) at (3.5,-1.9) {};
		\node[style=none] (xdot1l) at (2,-0.3) {};
		\node[style=boxsmall,draw=red] (xdot2ins) at (-3.5,-1.9) {};
		\node[style=boxsmall,draw=red] (xdot1ins) at (-3.5,-1.2) {};
		\node[style=none] (xdot1in) at (-5,-1.9) {};
		\node[style=none] (xdot2in) at (-5,-1.2) {};
	\end{pgfonlayer}
	\begin{pgfonlayer}{edgelayer}
		\draw[-] (idin) to (idout); 
		\draw[-,draw=red] (Uinl1) to (Uinl2); 
		\draw[-,draw=red,out=0,in=135] (Uinr1) to (mult); 
		\draw[-,draw=red] (Udagoutl1) to (Udagoutl2.center); 
		\draw[-,draw=red] (Udagoutr1) to (Udagoutr2.center); 
		\draw[-,draw=red,out=0,in=180] (mult) to (xdot1l); 
		\draw[-,draw=red,out=0,in=90] (Udagoutl2.center) to (xdot2); 
		\draw[-,draw=red,out=0,in=-90] (xdot2l.center) to (xdot2); 
		\draw[-,draw=red] (xdot2ins.center) to (xdot2l.center); 
		\draw[-,draw=red,out=0,in=-135] (xdot1ins.center) to (mult); 
		\draw[-,draw=red,out=0,in=180] (xdot2in.center) to (xdot2ins.center); 
		\draw[-,draw=red,out=0,in=180] (xdot1in.center) to (xdot1ins.center); 
	\end{pgfonlayer}

\end{tikzpicture}
	\hspace{3mm}= \hspace{3mm}
	\begin{tikzpicture}
	\begin{pgfonlayer}{nodelayer}
		\node[style=none] (in) at (0,+2.5) {};
		\node[style=none] (out) at (9,+2.5) {};
		\node[style=none,draw=red,text=red,minimum height=0mm] (iny) at (0,+1.25) {};
		\node[style=none,draw=red,text=red,minimum height=0mm] (iny0) at (0,0) {};
		\node[style=none,draw=red,text=red,minimum height=0mm] (inx) at (0,-1.25) {};
		\node[style=none,draw=red,text=red,minimum height=0mm] (inx0) at (0,-2.5) {};
		\node[style=none] (idy) at (2,+1.25) {};
		\node[style=none] (idy0) at (2,0) {};
		\node[style=boxsmall,draw=red] (antipodex) at (2,-1.25) {};
		\node[style=boxsmall,draw=red] (antipodex0) at (2,-2.5) {};
		\node[style=Zbwdot,draw=red] (posadd) at (4,0) {}; 
		\node[style=Zbwdot,draw=red] (timeadd) at (4,-1.25) {}; 
		\node[style=Zbwdot,draw=red] (unit) at (5,-0.625) {};
		\node[style=none] (Uinl) at (6.5,-0.625) {};
		\node[style=none] (Uinr) at (6.5,-1.25) {};
		\node[style=box,draw=red,text=red,minimum height=6mm,minimum width=6mm] (U) at (6.5,-0.9375) {$U$}; 
		\node[style=none] (Uout) at (6.5,-1.25) {};
		\node[style=none] (dotinl) at (6.5,0) {};
		\node[style=Zbwdot,draw=red] (dot) at (7.6,-0.625) {};
	\end{pgfonlayer}
	\begin{pgfonlayer}{edgelayer}
		\draw[-] (in) to (out);
		\draw[-,draw=red] (iny) to (idy.center);
		\draw[-,draw=red] (iny0) to (idy0.center);
		\draw[-,draw=red] (inx) to (antipodex);
		\draw[-,draw=red] (inx0) to (antipodex0);
		\draw[-,draw=red,out=0,in=+135] (idy.center) to (posadd);
		\draw[-,draw=red,out=0,in=+135] (idy0.center) to (timeadd);
		\draw[-,draw=red,out=0,in=-135] (antipodex) to (posadd);
		\draw[-,draw=red,out=0,in=-135] (antipodex0) to (timeadd);
		\draw[-,draw=red] (posadd) to (dotinl.center);
		\draw[-,draw=red] (unit) to (Uinl);
		\draw[-,draw=red] (timeadd) to (Uinr.center);
		\draw[-,draw=red,out=0,in=+90] (dotinl.center) to (dot);
		\draw[-,draw=red,out=0,in=-90] (Uinr.center) to (dot);

	\end{pgfonlayer}
\end{tikzpicture}
\end{align}
The first equation uses the definition of the Heisenberg picture operator $\bar{\gamma}^\dagger$. The second equation (up to $\tau$) uses the commutation relation for the operator $\gamma$. The third equation uses the snake equation for $\hbox{\begin{tikzpicture} [scale=1.2,transform shape] %% DO NOT CHANGE

\def\deltax{0.3} %% CAN BE CHANGED
\def\deltay{0.5} %% DO NOT CHANGE

%\path[use as bounding box] (-\deltax,-\deltay) rectangle (\deltax,\deltay);

\node [dot, fill=\Zbwcolour] (mult) at (0,0) {};
%\node (mult_label_out) at (0,+\deltay) {};
%\draw[-] (mult) to (mult_label_out);

%\draw (current bounding box.south west) rectangle (current bounding box.north east);
\end{tikzpicture}
}\!\!$. The fourth equation uses the unitariety condition for the module $U$. The fifth equation uses the multiplicativity condition for the module $U$. The sixth equation changes $\hbox{\begin{tikzpicture} [scale=1.2,transform shape] %% DO NOT CHANGE

\def\deltax{0.3} %% CAN BE CHANGED
\def\deltay{0.5} %% DO NOT CHANGE

%\path[use as bounding box] (-\deltax,-\deltay) rectangle (\deltax,\deltay);

\node [dot, fill=\Dcolour] (mult) at (0,0) {};
%\node (mult_label_out) at (0,+\deltay) {};
%\draw[-] (mult) to (mult_label_out);

%\draw (current bounding box.south west) rectangle (current bounding box.north east);
\end{tikzpicture}
}\!\!$ to $\hbox{\begin{tikzpicture} [scale=1.2,transform shape] %% DO NOT CHANGE

\def\deltax{0.3} %% CAN BE CHANGED
\def\deltay{0.5} %% DO NOT CHANGE

%\path[use as bounding box] (-\deltax,-\deltay) rectangle (\deltax,\deltay);

\node [dot, fill=\Zbwcolour] (mult) at (0,0) {};
%\node (mult_label_out) at (0,+\deltay) {};
%\draw[-] (mult) to (mult_label_out);

%\draw (current bounding box.south west) rectangle (current bounding box.north east);
\end{tikzpicture}
}\!\!$ by introducing an antipode. The seventh and last equation uses the fact that the unitary module is diagonalised by $\hbox{\begin{tikzpicture} [scale=1.2,transform shape] %% DO NOT CHANGE

\def\deltax{0.3} %% CAN BE CHANGED
\def\deltay{0.5} %% DO NOT CHANGE

%\path[use as bounding box] (-\deltax,-\deltay) rectangle (\deltax,\deltay);

\node [dot, fill=\Zbwcolour] (mult) at (0,0) {};
%\node (mult_label_out) at (0,+\deltay) {};
%\draw[-] (mult) to (mult_label_out);

%\draw (current bounding box.south west) rectangle (current bounding box.north east);
\end{tikzpicture}
}\!\!$ (and associativity of $\hbox{\begin{tikzpicture} [scale=1.2,transform shape] %% DO NOT CHANGE

\def\deltax{0.3} %% CAN BE CHANGED
\def\deltay{0.5} %% DO NOT CHANGE

%\path[use as bounding box] (-\deltax,-\deltay) rectangle (\deltax,\deltay);

\node [dot, fill=\Zbwcolour] (mult) at (0,0) {};
%\node (mult_label_out) at (0,+\deltay) {};
%\draw[-] (mult) to (mult_label_out);

%\draw (current bounding box.south west) rectangle (current bounding box.north east);
\end{tikzpicture}
}\!\!$).

\noindent We also do a step-by-step check that the propagator takes the form claimed in the body.
\begin{align}
	D(x-y)
	\hspace{2mm} =_\tau & \hspace{2mm}
	\begin{tikzpicture}
	\begin{pgfonlayer}{nodelayer}
		\node[style=none] (in) at (-2,+2.5) {};
		\node[style=none] (out) at (9,+2.5) {};
		\node[style=state,draw=red,text=red,minimum height=0mm] (iny) at (0,+1.25) {$\delta_{\underline{y}}$};
		\node[style=state,draw=red,text=red,minimum height=0mm] (iny0) at (0,0) {$\delta_{y_0}$};
		\node[style=state,draw=red,text=red,minimum height=0mm] (inx) at (0,-1.25) {$\delta_{\underline{x}}$};
		\node[style=state,draw=red,text=red,minimum height=0mm] (inx0) at (0,-2.5) {$\delta_{x_0}$};
		\node[style=none] (idy) at (2,+1.25) {};
		\node[style=none] (idy0) at (2,0) {};
		\node[style=boxsmall,draw=red] (antipodex) at (2,-1.25) {};
		\node[style=boxsmall,draw=red] (antipodex0) at (2,-2.5) {};
		\node[style=Zbwdot,draw=red] (posadd) at (4,0) {}; 
		\node[style=Zbwdot,draw=red] (timeadd) at (4,-1.25) {}; 
		\node[style=Zbwdot,draw=red] (unit) at (5,-0.625) {};
		\node[style=none] (Uinl) at (6.5,-0.625) {};
		\node[style=none] (Uinr) at (6.5,-1.25) {};
		\node[style=box,draw=red,text=red,minimum height=6mm,minimum width=6mm] (U) at (6.5,-0.9375) {$U$}; 
		\node[style=none] (Uout) at (6.5,-1.25) {};
		\node[style=none] (dotinl) at (6.5,0) {};
		\node[style=Zbwdot,draw=red] (dot) at (7.6,-0.625) {};
	\end{pgfonlayer}
	\begin{pgfonlayer}{edgelayer}
		\draw[-] (in) to (out);
		\draw[-,draw=red] (iny) to (idy.center);
		\draw[-,draw=red] (iny0) to (idy0.center);
		\draw[-,draw=red] (inx) to (antipodex);
		\draw[-,draw=red] (inx0) to (antipodex0);
		\draw[-,draw=red,out=0,in=+135] (idy.center) to (posadd);
		\draw[-,draw=red,out=0,in=+135] (idy0.center) to (timeadd);
		\draw[-,draw=red,out=0,in=-135] (antipodex) to (posadd);
		\draw[-,draw=red,out=0,in=-135] (antipodex0) to (timeadd);
		\draw[-,draw=red] (posadd) to (dotinl.center);
		\draw[-,draw=red] (unit) to (Uinl);
		\draw[-,draw=red] (timeadd) to (Uinr.center);
		\draw[-,draw=red,out=0,in=+90] (dotinl.center) to (dot);
		\draw[-,draw=red,out=0,in=-90] (Uinr.center) to (dot);

	\end{pgfonlayer}
\end{tikzpicture}
	\hspace{2mm} =
	\nonumber \\
	=&
	\hspace{2mm}
	\scalebox{0.95}{$\Bigg(
		\sum_{\underline{p}} \frac{1}{\omega_{ir}^3} \frac{1}{2E_{\underline{p}}} 
		\bra{\delta_{\underline{x}-\underline{y}}}U\big(\ket{\goodchi_{p}} \otimes \ket{\delta_{y_0-x_0}}\big)
	\Bigg) \id{}$}
	\hspace{2mm} = \hspace{2mm}
	\scalebox{0.95}{$\Bigg(
		\sum_{\underline{p}} \frac{1}{\omega_{ir}^3} \frac{1}{2E_{\underline{p}}} 
		e^{i2\pi E_{\underline{p}}\cdot(y_0-x_0)}
		\braket{\delta_{\underline{x}-\underline{y}}}{\goodchi_{p}}
	\Bigg) \id{}$}
	\nonumber \\
	=&
	\hspace{2mm}
	\scalebox{0.95}{$\Bigg(
		\sum_{\underline{p}} \frac{1}{\omega_{ir}^3} \frac{1}{2E_{\underline{p}}} 
		e^{-i2\pi \big[ E_{\underline{p}}\cdot(x_0-y_0) - \underline{p}\cdot (\underline{x}-\underline{y}) \big]}
		\braket{\delta_{\underline{x}-\underline{y}}}{\goodchi_{p}}
	\Bigg) \id{}$}
	\hspace{2mm} = \hspace{2mm}
	\scalebox{0.95}{$\Bigg(
		\sum_{\underline{p}} \frac{1}{\omega_{ir}^3} \frac{1}{2E_{\underline{p}}} 
		e^{-i2\pi p\cdot(x-y)} 
	\Bigg)\id{}$}
	\nonumber \\
\end{align}

% \newpage
% \section{Field-controlled creation/destruction operators}

% \noindent Still to be done:
% \begin{enumerate}
% 	\item coherently controlled creation/destruction operators are not algebraically nice: hard to characterise multiple creation/destruction steps
% 	\item what do we have that keeps track of numbers for each momentum value? The fields! 
% 	\item define the field-controlled creation/destruction operator
% 	\item classical structure for momentum eigenfields
% 	\item monoid for pointwise number addition, bialgebra laws
% 	\item field-controlled creation/destruction operators are modules
% 	\item monoid for pointwise number multiplication, modified bialgebra laws, distributivity
% 	\item commutation relations in terms of the field-controlled operators (and number operator)
% 	\item define the wavefunction-to-single-particle isometry using the vacuum and the coherently controlled operators
% 	\item single-particle projector from isometry
% 	\item number operator from field-controlled creation/destruction operators and single-particle projector
% 	\item hamiltonian?
% \end{enumerate}

\end{document}